\documentclass[11pt,a4paper]{article}
\usepackage{jcappub}
\usepackage{bm}
\usepackage{amsmath}
\def\dd{\mathrm{d}}
\def\mcA{\mathcal{A}}
\def\mcP{\mathcal{P}}
\def\ff{{\rm f}}
\def\em{{\rm em}}
\def\inf{{\rm inf}}
\def\tot{{\rm tot}}
\def\obs{{\rm obs}}
\def\mi{{\rm min}}
\def\ma{{\rm max}}
\def\Mpl{M_{\rm Pl}}
\def\GeV{{\rm GeV}}
\def\CMB{{\rm CMB}}

\title{
Higher order statistics of curvature perturbations
in $IFF$ model and its Planck constraints
}
\author[a,b]{Tomohiro Fujita}
\author[c]{Shuichiro Yokoyama}
\affiliation[a]{Kavli Institute for the Physics and Mathematics of the
Universe (Kavli IPMU), TODIAS,  the University of Tokyo, 5-1-5
Kashiwanoha, Kashiwa, 277-8583, Japan}
\affiliation[b]{Department of Physics, University of Tokyo, Bunkyo-ku
113-0033, Japan}
\affiliation[c]{Institute for Cosmic Ray Research, University of Tokyo,
5-1-5 Kashiwa-no-Ha, Kashiwa, Chiba, 277-8582, Japan}
\emailAdd{tomohiro.fujita@ipmu.jp}
\emailAdd{shu@icrr.u-tokyo.ac.jp}

\abstract{
We compute the power spectrum $\mathcal{P}_\zeta$ and
non-linear parameters $f_{NL}$ and $\tau_{NL}$ of the curvature
perturbation induced during inflation by the electromagnetic fields 
in the kinetic coupling model ($IFF$ model).  
By using the observational result of $\mathcal{P}_\zeta, f_{NL}$ and $\tau_{NL}$
reported by the Planck collaboration,
we study the constraint on the model comprehensively.
Interestingly, if the single slow-rolling inflaton is responsible for 
the observed $\mathcal{P}_\zeta$, the constraint from $\tau_{NL}$ is
most stringent.
We also find a general relationship between $f_{NL}$ and $\tau_{NL}$
generated in this model. Even if $f_{NL}\sim \mathcal{O}(1)$,
a detectable $\tau_{NL}$ can be produced.
}

\keywords{non-gaussianity, inflation, primordial magnetic fields}
\arxivnumber{1306.2992}

\begin{document}

\begin{flushright}
ICRR-Report-655-2013-4
\\
IPMU 13-0120
\end{flushright}

\maketitle

%
%
%
\section{Introduction}

Recently, a possibility of a vector field playing important roles
during inflation has been intensely studied. Although a U(1) gauge field
is not fluctuated during inflation in its minimal form due to 
the conformal symmetry,
several ideas to extend it are proposed.
Among them, the kinetic coupling model~\cite{Ratra:1991bn}
is nicely simple,
free of ghost instabilities~\cite{Himmetoglu:2008zp}
and well motivated by the supergravity 
or the string theory frame
work~\cite{Lemoine:1995vj,Gasperini:1995dh,Bamba:2003av,Bamba:2004cu,Ganjali:2005sr,Martin:2007ue}.
The model action is given by
\begin{equation}
S_A= \int \dd\eta \dd^3 x \sqrt{-g}
\left[
-\frac{1}{4} I^2(\phi) F_{\mu\nu}F^{\mu\nu}
\right],
\quad
\left(
F_{\mu\nu}\equiv \partial_\mu A_\nu - \partial_\nu A_\mu
\right),
\label{Model Action}
\end{equation}
where $A_\mu$ is a gauge field,
$\phi$ is a homogeneous and dynamical
scalar field which is not necessarily the inflaton
and $\eta$ is the conformal time.
Extensive literature explores its theoretical and observational consequences.

Earlier works are aimed at generating
the primordial magnetic field during inflation 
or ``inflationary magnetogenesis"~( e.g., \cite{Kandus:2010nw, Giovannini:2003yn} and reference therein).
It is observationally known that most galaxies and galaxy clusters 
have $\mathcal{O}(10^{-6})$G magnetic fields and recently
$\mathcal{O}(10^{-15})$G magnetic field in void regions are reported to be detected~\cite{Neronov:1900zz,Tavecchio:2010mk,Taylor:2011bn,Takahashi:2011ac}. 
Since no successful astrophysical mechanism which can illustrate their origins are known,
it is interesting to seek them in the inflation era.
Under such conditions,
the kinetic coupling model was expected to realize the magnetogenesis.
Unfortunately, however, it turns out that the model 
suffers from the so-called back reaction
problem~\cite{Bamba:2003av,Demozzi:2009fu, Kanno:2009ei,Fujita:2012rb}
to generate the primordial magnetic field enough to explain the observations.
The back reaction problem addresses
that the energy density of the 
electromagnetic fields should be less than the inflation energy density,
otherwise the consistency of inflationary magnetogenesis is invalid
 (see sec.~\ref{Review of the model}).
As another theoretical problem, so-called a strong coupling problem
 is also stressed~\cite{Demozzi:2009fu}. 
 This problem restricts 
 the small kinetic coupling $I(\phi) \ll 1$ during inflation to ensure
 the perturbative approach 
 in terms of quantum loop effects~\cite{Gasperini:1995dh,Demozzi:2009fu}.
\footnote{
However in ref.~\cite{Giovannini:2009xa} , the author claims
`` Since the inflationary
evolution commences in a regime of strong gravitational coupling, it is not unreasonable that
also the gauge coupling could be strong at the onset of the dynamical evolution"
and tolerates the strong coupling problem.
}
When these problems are taken seriously,
there does not exist any successful inflationary magnetogenesis scenario
even in the context of the kinetic coupling model.
\footnote{
While we were preparing this paper, ref.\cite{Ferreira:2013sqa}
appeared on the arXiv.
In ref.\cite{Ferreira:2013sqa}, the authors claimed a $10^{-16}$G magnetic field at present Mpc scale can be produced in the kinetic
coupling model if $I(\eta)$ is not a monotonic but a complicated function.
}
Beyond the context of the inflationary magnetogenesis to generate the observed magnetic fields,
recently, 
the gauge field has been focused on as 
a source of the adiabatic curvature perturbations and also the tensor perturbations
~\cite{Suyama:2012wh,Giovannini:2013rme,Sorbo:2011rz,Barnaby:2012xt}.
\footnote{
Ref.~\cite{Giovannini:2006ph,Giovannini:2006gz,Giovannini:2006kc,Giovannini:2007aq} are earlier intensive works. See also them.
}
It gives specific features in the perturbations, e.g., as 
a statistical anisotropy~\cite{Yokoyama:2008xw, Watanabe:2010fh,Emami:2011yi, Bartolo:2012sd,Emami:2013bk},
non-gaussianity~\cite{Karciauskas:2008bc,ValenzuelaToledo:2009nq,Dimastrogiovanni:2010sm, Barnaby:2012tk,Shiraishi:2013vja,Abolhasani:2013zya},
and cross correlations between the gauge field and the curvature/tensor perturbations
~\cite{Caldwell:2011ra,Motta:2012rn,Jain:2012ga}.
In other words,
in a similar way to the back reaction problem, it is expected that
the precise information about the primordial perturbations derived from 
the cosmological observations gives a new constraint on
the kinetic coupling model.
Quite recently, the Planck collaboration has reported updated observational
information about the primordial curvature perturbations, 
especially, e.g., the amplitude of 
curvature perturbation $\mathcal{P}_\zeta$, non-linearity  parameters $f_{\rm NL}$
and $\tau_{\rm NL}$ which represent the amplitudes of the bispectrum and
trispectrum respectively%
~\cite{Ade:2013zuv,Ade:2013ydc}.
Thus it is appropriate time 
to investigate the primordial curvature perturbations induced from the
gauge field in the kinetic coupling model precisely,
and to derive a constraint on the model.

In spite of its importance, limited attentions are paid
to induced curvature perturbations in the kinetic coupling model. 
Actually previous works are done only under either of following assumptions
\footnote{
See, however, ref.~\cite{Giovannini:2013rme} in which
the author calculates the power spectrum of induced $\zeta$
without these assumptions. But non-gaussianities are not computed there.
Ref.~\cite{Shiraishi:2012xt} also treats non-flat electromagnetic spectrum
while the generation of CMB temperature fluctuation after the end of inflation is studied.
}
;
(1) $I(\phi) \propto a^{\pm2}$ is given and it produces exact 
scale-invariant spectra of electric or magnetic fields.
(2) 
$\phi$ of $I(\phi)$ is the inflaton field, where
the quantum fluctuations of the inflaton is responsible for the dominant source of
the curvature perturbations and the effect of the gauge field on the inflaton fluctuations
through the direct coupling $I(\phi)F_{\mu\nu}F_{\mu\nu}$ is investigated.

In this paper, 
we consider more general situations, where
we 
specify neither the scalar field in the kinetic coupling, $I $,
nor the dominant source of the curvature perturbations 
and the functional form of $I$ is given by $I \propto a^{-n}$ for 
an arbitrary $n\ge 2$.
Our strategy is simple. We derive the evolution equation of $\zeta$
in the presence of electromagnetic fields and calculate its power spectrum $\mathcal{P}_\zeta$
and non-gaussianities ($f_{\rm NL}, \tau_{\rm NL})$
induced by electromagnetic field in the kinetic coupling model with $I(\phi)\propto a^{-n}$.
Then,  by using observation result of 
Planck collaboration~\cite{Ade:2013zuv,Ade:2013ydc},
we obtain the constraints on  the parameters of the model and inflation,
which are not only the tilt of electromagnetic fields spectrum
corresponding to the model parameter, $n$,
 but also inflation energy scale and total e-folding number. 
As a result, we find that 
the allowed
parameter region is reduced from the one where only the back reaction problem is
taken into account.
Interestingly, the constraint from $\tau_{\rm NL}$
is most stringent under the assumption that the dominant source of
the curvature perturbations is attributed to the quantum fluctuations of the inflaton field.
We also find that in the kinetic coupling model the large $\tau_{\rm NL}\ (\gtrsim 10^3)$ can be realized
even for the small $f_{\rm NL}\ (\lesssim 10)$.

The rest of paper is organized as follows.
In section \ref{Review of the model}, we review the kinetic coupling model
and discuss the back reaction problem.
In section \ref{Curvature perturbation induced by electromagnetic fields},
we derive the evolution equation of $\zeta$ induced by the electromagnetic
field during inflation. We also calculate its correlators up to 4-point
and obtain induced $\mathcal{P}_\zeta, f_{\rm NL}$ and $\tau_{\rm NL}$.
In section \ref{Observational constraints}, we compare these quantities 
to observational results and show the restricted parameter region.
We conclude in section \ref{Conclusion}.

\section{Review of the kinetic coupling model and the back reaction problem}
\label{Review of the model}

\subsection{Model set up}
\label{Model set up}

We consider the kinetic coupling 
model~\cite{Ratra:1991bn,Himmetoglu:2008zp,Bamba:2003av,Lemoine:1995vj,Gasperini:1995dh,Bamba:2004cu,Ganjali:2005sr,Martin:2007ue}
in this paper.
Although it can not generate the primordial magnetic field
which is strong enough to be more than $10^{-15}$G at 
present~\cite{Demozzi:2009fu, Kanno:2009ei,Fujita:2012rb},
it is nicely simple and gives us the essential understanding 
of the problem.
Moreover this model is interesting in terms of CMB observations because 
it can produce detectable level of non-gaussianities.
In this section we review the model.

In the kinetic coupling model, 
the kinetic term of U(1) gauge field is modified as
$F_{\mu\nu}F^{\mu\nu} \rightarrow I^2(\phi) F_{\mu\nu}F^{\mu\nu}$ 
where $\phi$ is a homogeneous scalar field and is not necessarily inflaton and $I(\phi)$ is 
phenomenologically assumed to be the power function of conformal time,
$I\propto \eta^n$. To restore the Maxwell theory after inflation,
$I$ is required to be unity at the end of inflation $\eta_\ff$.
Thus $I(\phi)$ is reduced as
\begin{equation}
I(\phi)=
\left\{
\begin{array}{cc}
 (\eta/\eta_\ff)^n &  (\eta<\eta_\ff)\\
 1 & (\eta\ge\eta_\ff)
\end{array}\,. \right.
\label{model action}
\end{equation}
We do not specify the Lagrangian of $\phi$ and assume the quasi de Sitter inflation, the Einstein gravity and the flat FLRW metric. 
Note that hereafter we consider only positive $n$ to avoid the strong coupling problem.
Because if $n$ is negative and the QED coupling $e\bar{\psi}\gamma^\mu \psi A_\mu$ exists, its effective coupling constant,
$e/I$, becomes much larger than unity during inflation. In that case, we can not calculate the behavior of $A_\mu$
without fully taking account of the interaction effects~\cite{Demozzi:2009fu}.

Let us take the radiation gauge, $A_0=\partial_iA_i=0$,
and expand the transverse part of $A_i$ with the polarization
vector $\epsilon_i^{(\lambda)}$ and the creation/annihilation operator 
$a^{\dagger(\lambda)}_{\bm{k}}/a^{(\lambda)}_{\bm{k}}$ as
\footnote{
The polarization vector $\epsilon_i^{(\lambda)}$ satisfies
$k_i \epsilon_{i}^{(\lambda)}(\hat{\bm{k}})=0,$ and
$\sum_{p=1}^{2} {\epsilon}_{i}^{(\lambda)}(\hat{\bm{k}}) {\epsilon}_{j}^{(\lambda)}(-\hat{\bm{k}})
=\delta_{ij} - (\hat{\bm{k}})_{i}(\hat{\bm{k}})_{j}$
and the creation/annihilation operators satisfy 
$[a^{(\lambda)}_{\bm{p}},a^{\dagger(\sigma)}_{-\bm{q}}]
= (2\pi)^3\delta(\bm{p}+\bm{q})\delta^{\lambda \sigma}$, as usual.}
\begin{equation}
 A_i(\eta, \bm{x})
 = 
 \sum_{\lambda=1}^{2} \int \frac{{\rm d}^3 k}{(2\pi)^3} 
 e^{i \bm{k \cdot x}} \epsilon_{i}^{(\lambda)}(\hat{\bm{k}}) 
 \left[ a_{\bm{k}}^{(\lambda)} \mcA_{k}(\eta) 
  + a_{-\bm{k}}^{\dag (\lambda)} \mcA_{k}^{*}(\eta) \right]
\,,
\label{introduction of creation/annihilation operator}
\end{equation}
where the hat of $\hat{\bm{k}}$ denotes the unit vector
and $(\lambda)$ is the polarization label. 
Notice the behavior of $\mathcal{A}_k$ does not depend on the polarization
in this model. 
The equation of motion during inflation is given by
\begin{equation}
\left[\partial_\eta^2 +k^2-\frac{n(n-1)}{\eta^2}\right](I\mcA_k)=0
\,.
\label{EoM of A}
\end{equation}
Assuming the Bunch-Davies vacuum, $I\mcA_k=(2k)^{-1/2}e^{ik\eta}$, 
in the sub-horizon limit, 
the asymptotic solution of eq.~(\ref{EoM of A}) in the super-horizon is
\begin{equation}
| I\mcA_k(\eta) | = 
\frac{\Gamma(n-1/2)}{\sqrt{2\pi k}}\left( \frac{-k\eta}{2}\right)^{1-n}
,
\quad \left(-k\eta \ll 1,\  n>\frac{1}{2} \right),
\label{Sol of A}
\end{equation}
where we have neglected the constant phase factor.
For $0<n<1/2$, the asymptotic solution is different
and the generated electromagnetic fields are weaker than the cases of $n>1/2$.
Hence we focus on $n>1/2$ hereafter.

At this point, we can acquire three important 
consequences
in this model.
First, the generated magnetic field is negligible compared with the
electric field. The power spectrum of electric and magnetic fields are given by
\begin{equation}
\mcP_E (\eta,k) \equiv \frac{k^3 |\partial_\eta \mcA_k|^2}{\pi^2 a^4},\qquad
\mcP_B (\eta,k) \equiv \frac{k^5 |\mcA_k|^2}{\pi^2 a^4},
\label{P of EB}
\end{equation}
where two polarization modes are already summed.
Then $\mcP_B/\mcP_E \simeq (-k\eta)^2$ and 
the magnetic field is much smaller than the electric field
in the super-horizon. Second, the unique model parameter $n$
controls both the time dependence and the tilt of the electromagnetic 
energy spectrum.
The energy contribution from each $\ln k$ mode of electric and magnetic fields
can be calculated from the action eq.~(\ref{Model Action})
,
\begin{align}
\frac{\dd \rho_E}{\dd \ln k}=
\frac{1}{2} I^2 \mcP_E(\eta, k)
=
\frac{\Gamma^2(n+\frac{1}{2})}{2^{2-2n}\pi^3} H^4
\left(-k\eta \right)^{2(2-n)},
\label{E and B spectrum}
\\
\frac{\dd \rho_B}{\dd \ln k}=
\frac{1}{2} I^2 \mcP_B(\eta, k)
=
\frac{\Gamma^2(n-\frac{1}{2})}{2^{4-2n}\pi^3} H^4
\left(-k\eta \right)^{2(3-n)},
\notag
\end{align}
where $H$ is Hubble parameter.
The above equation tells that the electric field grows (decays) and the
spectrum of the electric energy density is red-tilted (blue-tilted) for $n>2$ ($n<2$).
The flat spectrum can be realized in $n=2$ case where the electric field stays constant.
In the magnetic case, 
the border of $n$ is 3 in stead of 2.
Finally, the magnetic power spectrum at present is
\begin{equation}
\mcP^{1/2}_B (\eta_{\rm now},k)
=
\frac{\Gamma(n-\frac{1}{2})}{2^{\frac{3}{2}-n}\pi^{\frac{3}{2}}}
(a_\ff H)^{n-1} k^{3-n}
\sim
10^{23n-80} {\rm G}\times
\left( \frac{\rho_\inf^{1/4}}{10^{16}\GeV} \right)^{n-1}
\left( \frac{k}{1{\rm Mpc}^{-1}} \right)^{3-n},
\label{current B}
\end{equation}
where $\rho_\inf$ is the energy density of the inflaton
and $a_\ff$ is a scale factor at the end of inflation ($a=1$ at the present).
Here we assume the instant reheating and have $a_\ff = \rho_\gamma /\rho_\inf$
with $\rho_\gamma$ being the present energy density of the radiation which is given by
$\rho_\gamma\approx 5.7\times10^{-125}\Mpl^4$.
From the above expression, we find that 
$n \gtrsim 3$ is required
to make the cosmic magnetic field whose strength is more than
the observational lower bound from blazars,
$10^{-15}$G, at Mpc scale.

\subsection{back reaction problem}
\label{back reaction problem}

In sec.~\ref{Model set up}, we assume that inflation continues and the electromagnetic
generation does not change regardless of the amount of
the electromagnetic fields. 
But if the energy density of the electromagnetic field $\rho_\em$ becomes comparable with that of inflaton, 
inflation itself or the generation of electromagnetic fields must be altered.
 Thus for the consistency of the above calculation, $\rho_\em<\rho_\inf$
should be satisfied. Unfortunately, however, in the parameter range where
the generated magnetic field is enough strong to explain the blazar observation,
namely $n \gtrsim 3$, $\rho_\em$ becomes larger than $\rho_\inf$. This problem
is called  ``back reaction problem"
\footnote{In Ref. \cite{Kanno:2009ei}, the authors have investigated the possibility of the electromagnetic generation
by taking into account its back reaction and the dynamics of $\phi$. 
In their case, although the inflation still continues,
the generation of the electromagnetic field is altered and
fails to produce the magnetic field which is strong enough to explain the blazar observation.}
.

From eq.~(\ref{Sol of A}),
the energy density of electromagnetic field 
during inflation is given by
\begin{equation}
\rho_\em(\eta) \simeq \frac{I^2}{2} \int^{aH}_{k_\mi}\frac{\dd k}{k}
\mcP_E (\eta, k)
=
\frac{\Gamma^2(n+\frac{1}{2})}{2^{2-2n}\pi^3}
H^4
\left[\frac{\left(-k_\mi \eta\right)^{2(2-n)}-1}{2n-4}\right]
\,,
\end{equation}
where we ignore the contribution of $\mcP_B$ and
$k_\mi$ is the wave number of the mode which crosses the horizon when
$I(\eta)$ starts to behave as $(\eta/\eta_\ff)^n$.
Because of $-k_\mi \eta < 1$,
$\rho_\em(\eta)$ is an increasing function of $\eta$ for $n \ge 2 $
while the $\eta$ dependence is negligible for $n<2$.
Thus 
for $n \ge 2$,
it is sufficient to require $\rho_\inf > \rho_\em (\eta)$ at the end of inflation for its satisfaction over the entire period of inflation. 
This condition puts the upper limit on $\rho_\inf$,
\begin{equation}
\frac{\rho_\inf}{\Mpl^4}\, 
< \,
\frac{2^{2-2n} 3^2 \pi^3}{\Gamma^2(n+\frac{1}{2})}
D_n^{-1}(N_\tot)
\qquad
(n\ge2),
\label{backreaction rho upper bound}
\end{equation}
where $N_{\rm tot} \equiv -\ln \left|k_\mi \eta_\ff \right|$
and we define new function $D_n$ for later simplicity,\begin{equation}
D_n(X)\equiv
\frac{e^{(2n-4)X}-1}{2n-4},
\qquad
\lim_{n\rightarrow2} D_n(X) = X
\,.
\label{def of D}
\end{equation}
Substituting eq.~(\ref{backreaction rho upper bound})
into eq.~(\ref{current B}), one can obtain the upper limit
of the magnetic power spectrum at present. 
For example, the upper limits 
for $n=3$ are 
\begin{equation}
\mcP^{1/2}_B (\eta_{\rm now},k,n=3)
<
1.8\times 10^{-28}{\rm G} \times \exp\left[50-N_{\rm tot}\right].
\end{equation}
For $n>3$, the upper bound on $\mcP_B(\eta_{\rm now},k)$
is more stringent. 
Therefore the kinetic coupling model can not generate the primordial
magnetic field with 
sufficient strength because of 
the back reaction problem.

\section{Curvature perturbation induced by electromagnetic fields}
\label{Curvature perturbation induced by electromagnetic fields}

Recently the effect of  
vector
fields in the kinetic coupling model on the curvature perturbation draws attention. 
The electromagnetic fields behave as isocurvature perturbations and
they can source the adiabatic curvature perturbation on super-Hubble scales.
The induced curvature perturbation has distinguishing non-gaussianities
which can be large enough for detection~\cite{Barnaby:2012tk, Shiraishi:2013vja}.
Planck data released in this March has given precise information about
the primordial curvature perturbation and also tighter constraints on the
non-linearity parameters which parameterize the non-Gaussian features of the
primordial curvature perturbation.
These Planck constraints can translate into the limits on the parameters of the kinetic coupling model and inflation.
In this section, we derive the curvature perturbation induced by
the electromagnetic fields in the kinetic coupling model during inflation. Then we compute its two-point, three-point,  four-point correlators
and their related non-linearity parameters.

\subsection{Evolution equation of $\zeta_\em$}
\label{Evolution equation}

The curvature perturbation $\zeta(t,\bm{x})$ is defined as the perturbation of the scale factor $a(t,\bm{x})$
on the uniform density slice,
$\zeta (t,\bm{x}) \equiv \ln \left[ a(t,\bm{x})/a(t)\right]$ where $t$ is the cosmic time.
Let us derive the evolution equation of $\zeta(t,\bm{x})$.
The energy continuity equation holds
on super-Hubble scales~\cite{Lyth:2004gb},
\begin{align}
\dot{\rho}(t) 
&= -3 \frac{\dot{a}(t,\bm{x})}{a(t,\bm{x})} [\rho(t) + p(t,\bm{x})] \nonumber \\ 
&= -3\left(H(t)+\dot{\zeta}(t,\bm{x}) \right) \left[ \rho(t)+p(t)+\delta p_{\rm nad} (t,\bm{x}) \right] .
\end{align}
By subtracting its homogeneous part, we obtain the evolution equation of the curvature perturbation on super-Hubble scales,
\begin{equation}
\dot{\zeta}(t,\bm{x}) = -\frac{H(t) \delta p_{\rm nad}(t,\bm{x})}{\rho(t)+p(t)} \,.
\label{first EoM of zeta}
\end{equation}
Here the non-adiabatic pressure is 
defined as
$
\delta p_{\rm nad}(t,\bm{x}) \equiv \delta p(t,\bm{x}) -\frac{\dot{p}(t)}{\dot{\rho}(t)} \delta \rho(t,\bm{x}).
$
In our case 
where the background 
energy density is dominated by
the inflaton field and the energy density of the electromagnetic field is
treated as a perturbation,
we have
\begin{equation}
p_{\rm inf} \simeq -\left( 1 - \frac{2}{3} \epsilon \right) \rho_{\rm inf},\quad
\delta \rho_{\rm em} = 3 \delta p_{\rm em},
\end{equation}
where $\epsilon$ is the slow-roll parameter
and indices ``inf'' and ``em'' denote the contribution from inflaton and electromagnetic fields, respectively.
Hence eq.~(\ref{first EoM of zeta}) reads~\cite{Giovannini:2006gz,Giovannini:2007aq},
\begin{equation}
\dot{\zeta}^\em (t,\bm{x}) = - \frac{2H(t)}{\epsilon \rho_{\rm inf}} \delta \rho_{\rm em} (t,\bm{x}),
\end{equation}
in the leading order of $\epsilon$. 
Integrating it, we finally obtain the expression of curvature perturbation induced 
by electromagnetic fields as~\cite{Suyama:2012wh}
\begin{equation}
\zeta^\em(t,\bm{x}) = -\frac{2H}{\epsilon \rho_{\rm inf}} \int^t_{t_0} dt' \delta \rho_\em (t',\bm{x}),
\label{zeta}
\end{equation}
where $H, \epsilon$ and $\rho_{\rm inf}$ are assumed to be constant during inflation and $t_0$ denotes
an initial time when $\zeta^\em(t_0,\bm{x})=0$.
Let us assume that the electromagnetic fields are
originally absent before the generation during inflation and thus
all electromagnetic fields exist as perturbations,
and hence we have
$
\delta \rho_{\rm E} =
\rho_{\rm E} = \frac{1}{2} I^2(\eta) \bm{E}^2(\eta,\bm{x})
$
and neglect the contribution of the magnetic energy (see the discussion below eq.~(\ref{P of EB})).
By performing Fourier transformation of $\bm{E}(\eta,\bm{x})$,
the electromagnetic energy density 
is written in the convolution 
of two Fourier transformed electric fields as
\begin{equation}
\delta \rho_{\rm em}(\eta,\bm{k}) 
\simeq
\frac{1}{2} I^2(\eta) \iint \frac{\dd^3 p\,\dd^3 q}{(2\pi)^3}
\delta(\bm{p}+\bm{q}-\bm{k})
\bm{E}(\eta,\bm{p})\cdot \bm{E}(\eta,\bm{q})
\,.
\label{electric energy}
\end{equation}
By using eq.~(\ref{introduction of creation/annihilation operator}),
(\ref{Sol of A}), (\ref{electric energy}) and
the definition of the electric field,
$E_i\equiv a^{-2} \partial_\eta A_i$,
eq.~(\ref{zeta}) reads
\footnote{
To be precise, the constant phase of the mode function 
which is neglected in eq.~(\ref{Sol of A}) should be included
in eq.~(\ref{a+a expression}) like 
$\left( a^{(\lambda)}_{\bm{p}}e^{i\xi}+a^{\dagger(\lambda)}_{\bm{-p}}e^{-i\xi} \right)$  where $e^{i\xi}$ is the constant phase factor.
However, since such phase factors vanish after the calculation
of the vacuum expectation value, we suppress them.
}
\begin{multline}
\zeta^\em (\eta,\bm{k}) 
=
\frac{c^2_n\rho_\inf}{9\epsilon \Mpl^4}
\iint^{k_\ma}_{k_\mi} \frac{\dd^3 p\,\dd^3 q}{(2\pi)^3}\,
\delta(\bm{p}+\bm{q}-\bm{k})\,
p^{\frac{1}{2}-n} q^{\frac{1}{2}-n}
\\ \times
\sum_{\lambda,\sigma}
\epsilon^{(\lambda)}_i (\hat{\bm{p}})\, \epsilon^{(\sigma)}_i (\hat{\bm{q}})\,
\left( a^{(\lambda)}_{\bm{p}}+a^{\dagger(\lambda)}_{\bm{-p}} \right)
\left( a^{(\sigma)}_{\bm{q}}+a^{\dagger(\sigma)}_{\bm{-q}} \right)
\int^\eta_{\eta_0}\dd\tilde{\eta}\, \tilde{\eta}^{3-2n}
\label{a+a expression}
\end{multline}
where the lower end of the time integration, $\eta_0=-\ma[p, q]^{-1}$,
represents that only super-horizon modes
are considered as physical modes,
$k_\ma = -\eta_\ff^{-1}$ is the maximum wave number
exiting the horizon during inflation and we define $c_n$ as
\begin{equation}
I\partial_\eta A_k(\eta)=c_n\, k^{\frac{1}{2}-n} \eta^{-n},
\quad
c_n\equiv\frac{2^n \Gamma(n+\frac{1}{2})}{\sqrt{2\pi}}
\,.
\end{equation}

Before closing this subsection,
let us note that the anisotropic stress which can also
source the curvature perturbation is not %
taken into account here.
However,
the contribution from the electromagnetic anisotropic stress
is suppressed by slow-roll parameter $\epsilon$ in comparison to
the contribution from the non-adiabatic pressure during inflation~\cite{Suyama:2012wh}.
Thus eq.~(\ref{zeta}) is the leading order equation.

\subsection{Calculation of 2, 3, 4-point correlators}
\label{Calculation of power spectrum}

Let us calculate two, three and four-point correlation function of the curvature perturbation in the Fourier space.
At first, we consider m-point correlator,
\begin{multline}
\left\langle 
\prod_{i=1}^m\zeta^\em(\eta,\bm{k}_i)
\right\rangle
=
\left\langle
\prod_{i=1}^m 
\left(\frac{c^2_n\rho_\inf}{9\epsilon \Mpl^4}\right)
\iint^{k_\ma}_{k_\mi} \frac{\dd^3 p_i\,\dd^3 q_i}{(2\pi)^3}
\delta(\bm{p}_i+\bm{q}_i-\bm{k}_i)
p_i^{\frac{1}{2}-n} q_i^{\frac{1}{2}-n}
\right.\\ \left.
\times
\sum_{\lambda_i,\sigma_i}
\epsilon^{(\lambda_i)}_{j_i}(\hat{\bm{p}_i}) \epsilon^{(\sigma_i)}_{j_i}(\hat{\bm{q}_i})
\left( a^{(\lambda_i)}_{\bm{p}_i}+a^{\dagger(\lambda_i)}_{\bm{-p}_i} \right)
\left( a^{(\sigma_i)}_{\bm{q}_i}+a^{\dagger(\sigma_i)}_{\bm{-q}_i} \right)
\int^\eta_{\eta_{0,i}}
\dd\tilde{\eta}_i\,
\tilde{\eta}_i^{3-2n}
\right\rangle
\,,
\label{m-point expression}
\end{multline}
where the bracket $\langle\cdots\rangle$ denotes the vacuum expectation value
and is only relevant to $a_{\bm{k}}^{(\lambda)}$ and 
$a_{-\bm{k}}^{\dagger(\lambda)}$.
One can show $\left\langle m{\rm -point}\right\rangle \equiv
\left\langle
\prod_{i=1}^m
\left( a^{(\lambda_i)}_{\bm{p}_i}+a^{\dagger(\lambda_i)}_{\bm{-p}_i} \right)
\left( a^{(\sigma_i)}_{\bm{q}_i}+a^{\dagger(\sigma_i)}_{\bm{-q}_i} \right)
\right\rangle
$
is given by
\begin{align}
&\left\langle 2{\rm -point}\right\rangle
=2(2\pi)^6 \delta(\bm{p}_1+\bm{q}_2) \delta(\bm{p}_2+\bm{q}_1)
\delta^{\lambda_1 \sigma_2} \delta^{\lambda_2 \sigma_1}
\,,
\label{a culc of 2-pint}
\\
&\left\langle 3{\rm -point}\right\rangle
=8(2\pi)^9 \delta(\bm{p}_1+\bm{q}_2) \delta(\bm{p}_2+\bm{q}_3)
\delta(\bm{p}_3+\bm{q}_1)
\delta^{\lambda_1 \sigma_2} \delta^{\lambda_2 \sigma_3} 
\delta^{\lambda_3 \sigma_1}
\,,
\label{a culc of 3-pint}
\\
&\left\langle 4{\rm -point}\right\rangle
=16\left\{
(2\pi)^{12} \delta(\bm{p}_1+\bm{q}_2) \delta(\bm{p}_2+\bm{q}_3)
\delta(\bm{p}_3+\bm{q}_4) \delta(\bm{p}_4+\bm{q}_1)
\delta^{\lambda_1 \sigma_2} \delta^{\lambda_2 \sigma_3} 
\delta^{\lambda_3 \sigma_4} \delta^{\lambda_4 \sigma_1}
\right.\notag\\ 
&\qquad\qquad\qquad\quad
\left. +(2 \leftrightarrow 3)+(3 \leftrightarrow 4) \Big\}
+({\rm disconnected \ terms})\right.
\,,
\label{a culc of 4-pint}
\end{align}
Since the calculation processes for $m=$2, 3 and 4 are analogous,
we illustrate only the $m=2$ case in detail.
By virtue of the delta function and the Kronecker delta in eq.~(\ref{a culc of 2-pint}),
the polarization factor in eq.~(\ref{m-point expression}) reads
\begin{equation}
\sum_{\lambda_1, \lambda_2}
\epsilon^{(\lambda_1)}_{j_1}(\hat{\bm{p}_1}) \epsilon^{(\lambda_2)}_{j_1}(-\hat{\bm{p}_2})
\epsilon^{(\lambda_2)}_{j_2}(\hat{\bm{p}_2}) \epsilon^{(\lambda_1)}_{j_2}(-\hat{\bm{p}_1})
=
\Bigl(\delta_{j_{1} j_{2}}-(\hat{\bm{p}}_1)_{j_1}(\hat{\bm{p}}_1)_{j_2}\Bigr)
\Bigl(\delta_{j_{1} j_{2}}-(\hat{\bm{p}}_2)_{j_1}(\hat{\bm{p}}_2)_{j_2}\Bigr)
\,.
\end{equation}
and the $\tilde{\eta}$ integral in eq.~(\ref{m-point expression}) reads
\begin{equation}
\prod_{i=1}^2
\int^\eta_{\eta_0}
\dd\tilde{\eta}_i\,
\tilde{\eta}_i^{3-2n}
=
\left[ \frac{\eta^{4-2n}-(-\ma[p_1,p_2])^{2n-4}}{2n-4}\right]^2
\,.
\end{equation}
Next one can perform the $q_i$ integrals by using $\delta(\bm{p}_i+\bm{q}_{i+1})$.
In the $m=2$ case, we obtain
\begin{multline}
\left\langle 
\zeta^\em_{\bm{k}_1} \zeta^\em_{\bm{k}_2} (\eta)
\right\rangle
=
2\delta(\bm{k}_1+\bm{k}_2) \left(\frac{c^2_n\rho_\inf}{9\epsilon \Mpl^4}\right)^2
\iint^{k_\ma}_{k_\mi} \dd^3 p_1 \dd^3 p_2
\delta(\bm{p}_2-\bm{p}_1-\bm{k}_2)\,
p_1^{1-2n}p_2^{1-2n}
\\ \times
\Bigl(\delta_{j_{1} j_{2}}-(\hat{\bm{p}}_1)_{j_1}(\hat{\bm{p}}_1)_{j_2}\Bigr)
\Bigl(\delta_{j_{1} j_{2}}-(\hat{\bm{p}}_2)_{j_1}(\hat{\bm{p}}_2)_{j_2}\Bigr)
\left[ \frac{\eta^{4-2n}-(-\ma[p_1,p_2])^{2n-4}}{2n-4}\right]^2
\,.
\label{Pole integral}
\end{multline}
If $n\ge2$, the biggest contributions of the integrals in eq.~(\ref{Pole integral})
come from the pole where $p_1\simeq k_\mi$ and $p_2\simeq k_\mi$. In the rest of this paper, we concentrate on the cases where $n\ge2$.
Then eq.~(\ref{Pole integral}) can be evaluated by the pole contributions.
Note the integrand has the symmetry of $\bm{p}_1 \leftrightarrow \bm{p}_2$.
Even in the case of $m=3$ and 4, the cyclic symmetry like,
$\bm{p}_1\rightarrow \bm{p}_2 \rightarrow  \cdots \rightarrow \bm{p}_m\rightarrow \bm{p}_1$,
exists. Thus if the $p_1$ pole is evaluated, the other contributions can be easily duplicated. The $p_1$ pole contribution in eq.~(\ref{Pole integral})
is evaluated as
\begin{equation}
\left\langle 
\zeta^\em_{\bm{k}_1} \zeta^\em_{\bm{k}_2} (\eta)
\right\rangle \big|_{p_1 \simeq k_\mi}
=
\frac{32\pi}{3\,k_1^3}\delta(\bm{k}_1+\bm{k}_2) \left(\frac{c^2_n\rho_\inf}{9\epsilon \Mpl^4}\right)^2
\left[\frac{(k_1/k_\mi)^{2n-4}-1}{2n-4} \right]
\left[\frac{(-k_1 \eta)^{4-2n}-1}{2n-4} \right]^2 ,
\end{equation}
where we use the angular integral,
$\int \dd \Omega_k \hat{\bm{k}}_i \hat{\bm{k}}_j = \frac{4\pi}{3}\delta_{ij}$,
and assume $k_1 = k_2 \gg k_\mi$.
\footnote{The assumption of $k_\CMB \gg k_\mi$ which corresponds to $N_\CMB < N_\tot$ means the generation of electromagnetic
fields begins much {\it earlier} than the horizon-crossing of CMB modes.
Although it may be interesting to consider the case where it begins {\it after} the CMB scale horizon-crossing, we focus on the former case in this paper.}
Notice additional factors like $(\ma[k_1,k_3]/\mi[k_1,k_3])^{2n-4}\ge1$ appear
in the case of $m=3$ and 4. Nevertheless,
we conservatively ignore those factors for simplicity 
by assuming all reference wave numbers are close to the CMB scale,
$k_i \sim k_\CMB$. 
Except for this point, the calculations of $m=3,4$ case
are closely analogous to $m=2$ case.
Therefore we obtain 2, 3 and 4-point connected correlation function of the electromagnetic
induced curvature perturbation at the end of inflation $\eta_\ff$ as
\begin{align}
\left\langle 
\zeta^\em_{\bm{k}_1} \zeta^\em_{\bm{k}_2} (\eta_\ff)
\right\rangle 
&=
\frac{64\pi}{3k_1^3}\delta(\bm{k}_1+\bm{k}_2) 
\left(\frac{c^2_n\rho_\inf}{9\epsilon \Mpl^4}\right)^2
D_n(N_\tot-N_\CMB) D_n(N_\CMB)^2 ,
\label{2-point correlator}
\\
\left\langle 
\zeta^\em_{\bm{k}_1} \zeta^\em_{\bm{k}_2} \zeta^\em_{\bm{k}_3}
(\eta_\ff)
\right\rangle 
&=
\frac{64\pi}{3}\delta(\bm{k}_1+\bm{k}_2+\bm{k}_3) 
\left(\frac{c^2_n\rho_\inf}{9\epsilon \Mpl^4}\right)^3
D_n(N_\tot-N_\CMB) D_n(N_\CMB)^3
\notag
\\
&\qquad \times
\left[ \frac{1+(\hat{\bm{k}}_1\cdot\hat{\bm{k}}_2)^2}{(k_1 k_2)^3}
+2\ {\rm perms} \right],
\label{3-point correlator}
\\
\left\langle 
\zeta^\em_{\bm{k}_1} \zeta^\em_{\bm{k}_2} \zeta^\em_{\bm{k}_3} \zeta^\em_{\bm{k}_4}
(\eta_\ff)
\right\rangle 
&=
\frac{128\pi}{3}\delta(\bm{k}_1+\bm{k}_2+\bm{k}_3+\bm{k}_4) 
\left(\frac{c^2_n\rho_\inf}{9\epsilon \Mpl^4}\right)^4
D_n(N_\tot-N_\CMB) D_n(N_\CMB)^4
\notag \\
&\hspace{-3cm}\times
\left[ \frac{
(\hat{\bm{k}}_1\cdot\hat{\bm{k}}_2)^2+(\hat{\bm{k}}_1\cdot\hat{\bm{k}}_{13})^2
+(\hat{\bm{k}}_2\cdot\hat{\bm{k}}_{13})^2
-(\hat{\bm{k}}_1\cdot\hat{\bm{k}}_2)(\hat{\bm{k}}_1\cdot\hat{\bm{k}}_{13})(\hat{\bm{k}}_2\cdot\hat{\bm{k}}_{13})
}{(k_1 k_2 k_{13})^3}
+11\ {\rm perms} \right],
\label{4-point correlator}
\end{align} 
where $\bm{k}_{13}\equiv \bm{k}_1 +\bm{k}_3\,,
D_n(X)\equiv (e^{(2n-4)X}-1)/(2n-4)\,,
e^{-N_\CMB} = -k_\CMB \eta_\ff \,$
and $e^{N_\tot-N_\CMB} = k_\CMB/k_\mi\,.$
In the limit of $n \rightarrow 2$, these results coincide with the previous
works~\cite{Barnaby:2012tk,Shiraishi:2013vja}.

When $n<2$, the correlators of induced $\zeta$ can not be computed as above because there is no pole.
Then we have to calculate the correlators by brute force.
But if $n$ is not too close to 2, the results are expected to depend on
neither $N_\tot$ nor $N_\CMB$. It is because the source of curvature perturbation,
$I^2\mathcal{P}_E(\eta,k)$, drops in the super-horizon as $\eta^{2(2-n)}$
and thus it sources $\zeta$ right after its horizon-crossing only.
Therefore since the resultant correlators are not just
much weaker than those in $n\ge2$ case but depend on neither $N_\tot$ nor $N_\CMB$, the motivation to constrain them is inadequate.
In this paper, we concentrate on the cases where $n\ge2$.

\subsection{Power spectrum and Non-gaussianities}

Let us connect 2,3,4-point correlators to the observable quantities
in order to compare them with the CMB observation results.
Here relevant observable quantities are the power spectrum of the primordial curvature perturbations
$\mathcal{P}_\zeta$,
and local-type non-linearity parameters $f_{\rm NL}^{\rm local}$ and $\tau_{\rm NL}$
which parameterize the amplitudes of the 3- and 4-point functions of the curvature perturbations in Fourier space, respectively.
These are defined as 
\begin{align}
\left\langle 
\zeta_{\bm{k}_1} \zeta_{\bm{k}_2}
\right\rangle
&=
(2\pi)^3\delta(\bm{k}_1+\bm{k}_2)
\frac{2\pi^2}{k_1^3}
\mathcal{P}_\zeta\,,
\label{def of power spectrum}
\\
\left\langle 
\zeta_{\bm{k}_1} \zeta_{\bm{k}_2} \zeta_{\bm{k}_3}
\right\rangle
&=
(2\pi)^3\delta(\bm{k}_1+\bm{k}_2+\bm{k}_3)\,
(2\pi^2\, \mathcal{P}_\zeta)^2\
\frac{6}{5}\, f_{\rm NL}^{\rm local}\,
\frac{\sum_{i=1}^3 k_i^3}{\prod_{i=1}^3 k_i^3}\,,
\label{def of fnl}
\\
\left\langle 
\zeta_{\bm{k}_1} \zeta_{\bm{k}_2} \zeta_{\bm{k}_3} \zeta_{\bm{k}_4}
\right\rangle
&=
(2\pi)^3\delta(\bm{k}_1+\bm{k}_2+\bm{k}_3+\bm{k}_4)\,
(2\pi^2 \mathcal{P}_\zeta)^3\,
\tau_{\rm NL}
\notag \\
&\hspace{4cm}\times\left[ \frac{1}{(k_1 k_2 k_{13})^3} + 11\ {\rm perms}
\right],
\label{def of taunl}
\end{align}
where the small deviation from scale invariant spectrum of $\mathcal{P}_\zeta$
is neglected.
By substituting eq.~(\ref{2-point correlator}) into eq.~(\ref{def of power spectrum}), one can easily obtain the induced power spectrum as
\begin{equation}
\mathcal{P}_\zeta^\em (k,\eta_\ff)
=
\frac{4}{3} 
\left(\frac{c^2_n\rho_\inf}{9\pi^2\epsilon \Mpl^4}\right)^2
D_n(N_\tot-N_\CMB) D_n^2(N_\CMB) .
\label{em power spectrum}
\end{equation}
As for $f_{\rm NL}^{\rm local}$ and $\tau_{\rm NL}$, however, 
$k_i$ dependence of eq.~(\ref{3-point correlator}) and
(\ref{4-point correlator}) is different from 
that of eq.~(\ref{def of fnl}) and (\ref{def of taunl}), respectively.
Thus they can not be compared straightforwardly
\footnote{Planck team also investigated the bispectrum which has such non-trivial $k_i$ dependences \cite{Ade:2013ydc}. In order to parameterize the angular dependence of the bispectrum they introduced the Legendre Polynomial expansion \cite{Shiraishi:2013vja}, and they obtained the constraint on each coefficient of the expansion. The result seems to be almost comparable to the constraint on $f_{\rm NL}^{\rm local}$ and hence for simplicity we apply $f_{\rm NL}^{\rm local}$ constraint to our result.}.

But when eq.~(\ref{3-point correlator}) and
(\ref{4-point correlator}) are averaged over the direction of $\hat{\bm{k}}_i$,
their $k_i$ dependence accord with that of eq.~(\ref{def of fnl}) and (\ref{def of taunl}), respectively.
\footnote{Taking angular average, one can show 
$(\hat{\bm{k}}_1\cdot \hat{\bm{k}}_2)^2$ yields 1/3 if these two unit vectors
are independent. But for example, the averaged value of 
$(\hat{\bm{k}}_1\cdot \hat{\bm{k}}_{13})^2$
depends on $k_1$ and $k_3$.
In the limit of $k_1 = k_3$, which is the squeezed limit where the terms with $\bm{k}_{13}$ become most important, the averaged $(\hat{\bm{k}}_1\cdot \hat{\bm{k}}_{13})^2$ is 1/2 and averaged 
$(\hat{\bm{k}}_1\cdot\hat{\bm{k}}_2)(\hat{\bm{k}}_1\cdot\hat{\bm{k}}_{13})(\hat{\bm{k}}_2\cdot\hat{\bm{k}}_{13})$
is 1/6. Thus we approximate the angular averaged value of the product of vectors depending each other by that in the relevant squeezed limit.
}
After angular averaged, eq.~(\ref{3-point correlator}) and
(\ref{4-point correlator}) read
\begin{align}
\left\langle 
\zeta^\em_{\bm{k}_1} \zeta^\em_{\bm{k}_2} \zeta^\em_{\bm{k}_3}
(\eta_\ff)
\right\rangle_{\rm ave} 
=
\frac{2^8\pi}{3^2}\delta(\bm{k}_1+\bm{k}_2+\bm{k}_3) 
\left(\frac{c^2_n\rho_\inf}{9\epsilon \Mpl^4}\right)^3
D_n(N_\tot-N_\CMB) D_n^3(N_\CMB)
\frac{\sum_{i=1}^3 k_i^3}{\prod_{i=1}^3 k_i^3}
\\
\left\langle 
\zeta^\em_{\bm{k}_1} \zeta^\em_{\bm{k}_2} \zeta^\em_{\bm{k}_3} \zeta^\em_{\bm{k}_4}
(\eta_\ff)
\right\rangle_{\rm ave}
=
\frac{2^{7}\pi}{3}\delta(\bm{k}_1+\bm{k}_2+\bm{k}_3+\bm{k}_4) 
\left(\frac{c^2_n\rho_\inf}{9\epsilon \Mpl^4}\right)^4
D_n(N_\tot-N_\CMB) D_n^4(N_\CMB)
\notag \\
\times\left[ \frac{1}{(k_1 k_2 k_{13})^3} + 11\ {\rm perms}
\right].
\label{averaged 4-point correlator}
\end{align} 
Therefore we obtain electromagnetic induced local-type non-gaussianities
\begin{align}
&f_{\rm NL}^\em
=
\frac{2^2 5}{3^3}
\left(\frac{c^2_n\rho_\inf}{9\pi^2\epsilon \Mpl^4}\right)^3
\mathcal{P}_\zeta^{-2}\,
D_n(N_\tot-N_\CMB) D_n^3(N_\CMB),
\label{em fnl}
\\
&\tau_{\rm NL}^\em =
\frac{2}{3}
\left(\frac{c^2_n\rho_\inf}{9\pi^2\epsilon \Mpl^4}\right)^4
\mathcal{P}_\zeta^{-3}\,
D_n(N_\tot-N_\CMB) D_n^4(N_\CMB).
\label{em taunl}
\end{align}
Note our three results can be written in the similar form as
\begin{align}
\mathcal{P}_\zeta^\em \simeq D_n(N_\tot-N_\CMB)\,G_n^2,
\label{abb form P}
\\
f_{\rm NL}^\em \mathcal{P}_\zeta^2 \simeq D_n(N_\tot-N_\CMB)\,G_n^3,
\\
\tau_{\rm NL}^\em \mathcal{P}_\zeta^3\simeq D_n(N_\tot-N_\CMB)\,G_n^4,
\label{abb form tau}
\end{align}
where $G_n \equiv c_n^2 \rho_\inf D_n(N_\CMB)/9\pi^2 \epsilon \Mpl^4$
and $\mathcal{O}(1)$ numerical factors are dropped. 
Then we obtain the general relationship between $f_{\rm NL}^\em$ and $\tau_{\rm NL}^\em$ in the kinetic coupling model of $n\ge2$,
\begin{equation}
\tau^\em_{\rm NL} \simeq 
\left[
\mathcal{P}_\zeta D_n(N_\tot -N_\CMB)
\right]^{-\frac{1}{3}}
{f_{\rm NL}^\em}^{\frac{4}{3}}\,.
\label{f and t relationship}
\end{equation}
Therefore even if $f_{\rm NL} \sim \mathcal{O}(1)$,
the kinetic coupling model can produce a large $\tau_{\rm NL}$.
\section{Observational constraints}
\label{Observational constraints}

In this section, we translate the Planck constraints 
on $\mathcal{P}_\zeta$, $f_{\rm NL}^{\rm local}$ and $\tau_{\rm NL}$ \cite{Ade:2013zuv,Ade:2013ydc}
into the constraints on the model parameters of kinetic coupling model. Planck collaboration reports:
\begin{align}
&\mathcal{P}_\zeta (k_\CMB)  \approx 2.2\times10^{-9}\,,
\label{Planck power spectrum}
\\
&f_{\rm NL}^{\rm local} \le \ f_{\rm NL}^\obs\equiv 14.3 \quad   (95\% {\rm CL})\,,
\\
&\tau_{\rm NL}\ \le \ \tau_{\rm NL}^\obs\equiv 2800 \quad   (95\% {\rm CL})\,.
\end{align}
The expressions of these observable quantities predicted in
the kinetic coupling model, namely eq.~(\ref{em power spectrum}),
(\ref{em fnl}) and (\ref{em taunl}), include four unknown parameters
$n, \epsilon, N_\tot$ and $\rho_\inf$.
Therefore, when three parameters out of four are fixed,
the other one can be constrained by the observation. 
Note that $N_\CMB$ can be estimated as
\begin{equation}
N_\CMB \simeq 62\ + \ln\left(\frac{\rho_\inf^{1/4}}{10^{16}\GeV}\right),
\label{NCMB}
\end{equation}
where the instantaneous reheating is assumed for simplicity.
In addition, if one assume the dominant component of the power spectrum of the curvature perturbation
is generated by a single slow-rolling inflaton, 
the curvature perturbation, $\mathcal{P}_\zeta$, is given by
\begin{equation}
\mathcal{P}^\inf_\zeta \equiv \frac{\rho_\inf}{24\pi^2 \epsilon \Mpl^4},
\label{slow-rolling inflaton}
\end{equation}
and then $\epsilon$ can be determined by $\rho_\inf$ under eq.~(\ref{Planck power spectrum}).
However, this assumption is not mandatory because the dominant component of the curvature perturbation
can be generated by the other mechanism like curvaton or modulated reheating\footnote{
Here, we neglect the non-Gaussianity generated in the curvaton or modulated reheating scenario.}.
Let us call the $\mathcal{P}_\zeta=\mathcal{P}^\inf_\zeta$ case ``inflaton" case while the conservative case where $\mathcal{P}_\zeta=\mathcal{P}^\inf_\zeta$ is not assumed is called ``curvaton" case although we do not specify the generation mechanism of $\mathcal{P}_\zeta$ as curvaton or any other models.

\subsection{Constraint on $N_\tot-N_\CMB$}

First, let us discuss the constraint on $N_\tot - N_\CMB$ with changing the parameter $n$.
Since we assume $N_\tot > N_\CMB$ in the derivation of eq.~(\ref{em power spectrum}), (\ref{em fnl}) and (\ref{em taunl}), we only consider 
the positive value of $N_\tot - N_\CMB$ for consistency.
Combined with the restriction that 
$\mathcal{P}_\zeta^\em, f^{\rm local}_{\rm NL}$ and $\tau_{\rm NL}^\em$
can not exceed the observed value or upper limits,
eq.~(\ref{backreaction rho upper bound}), (\ref{em power spectrum}),
(\ref{em fnl}) and (\ref{em taunl}) can be rewritten as
\begin{align}
&{\rm BR}:
&N_\tot-N_\CMB <&\ \frac{1}{2n-4}\ln\left[
1+(n-2) \left( \frac{6\pi}{c_n}\right)^2
\frac{\Mpl^4}{\rho_\inf}
\right]-N_\CMB,
\label{NtNc from BR}
\\
&\mathcal{P}_\zeta:
&N_\tot-N_\CMB \le &\ \frac{1}{2n-4}\ln\left[
1+ (n-2)\,\frac{3}{2}\, \, \mathcal{P}_\zeta \, G_n^{-2}
\right],
\\
&f_{\rm NL}^{\rm local}:
&N_\tot-N_\CMB \le &\ \frac{1}{2n-4}\ln\left[
1+ (n-2)\,\frac{27}{10}\   f_{\rm NL}^\obs\,\mathcal{P}_\zeta^2 \, G_n^{-3}
\right],
\\
&\tau_{\rm NL}:
&N_\tot-N_\CMB \le &\ \frac{1}{2n-4}\ln\left[
1+ (n-2)\,\frac{27}{8}\ \tau_{\rm NL}^\obs\,\mathcal{P}_\zeta^3 \, G_n^{-4}
\right],
\label{NtNc from taunl}
\end{align}
where 
$G_n  = \frac{8}{3} c_n^2 \mathcal{P}_\inf D_n(N_\CMB)$ by using eq. (\ref{slow-rolling inflaton})
and ``BR" denotes the constraint from the back reaction problem.

\begin{figure}[htbp]
  \hspace{-2mm}
  \includegraphics[width=75mm]{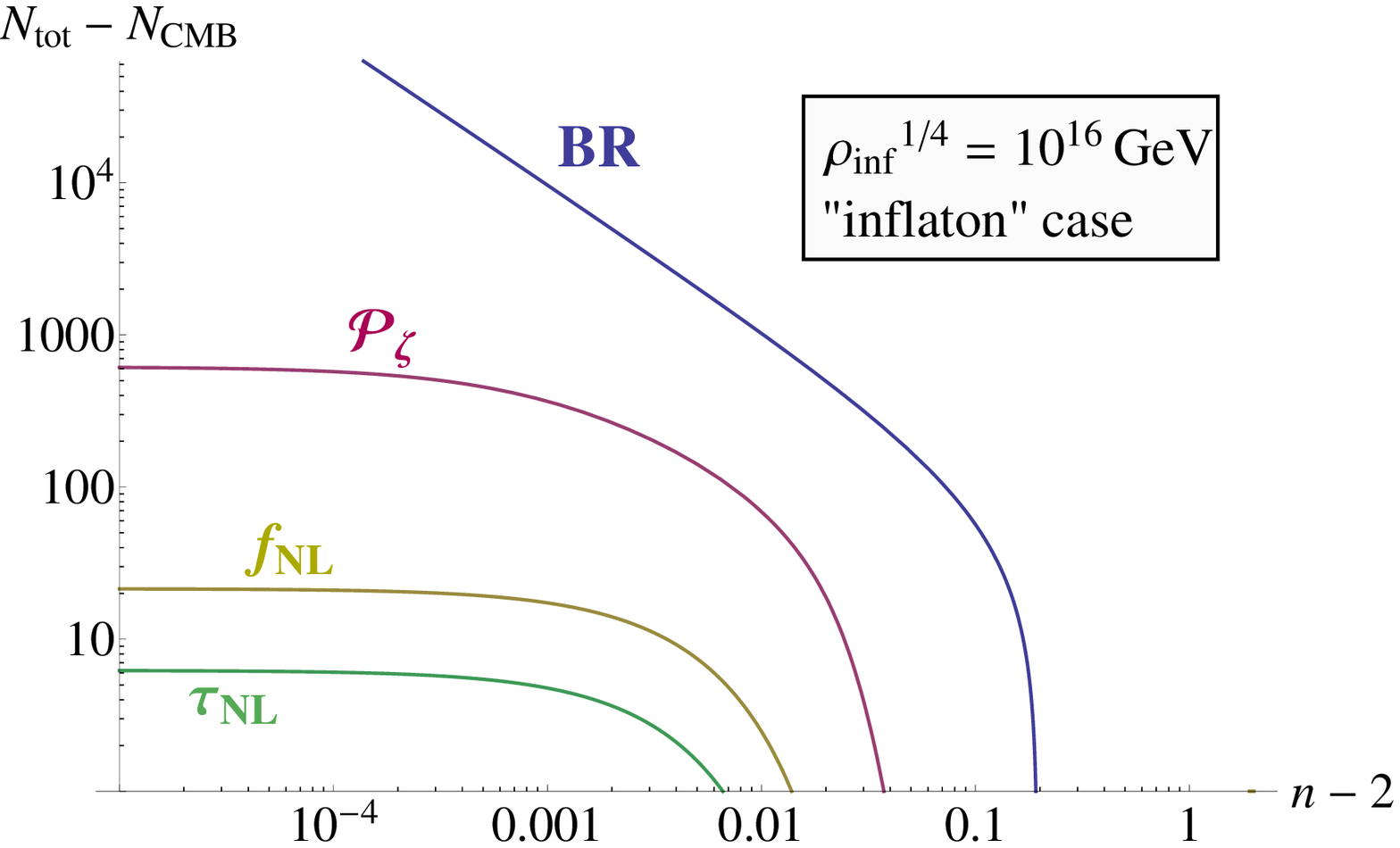}
  \hspace{5mm}
  \includegraphics[width=75mm]{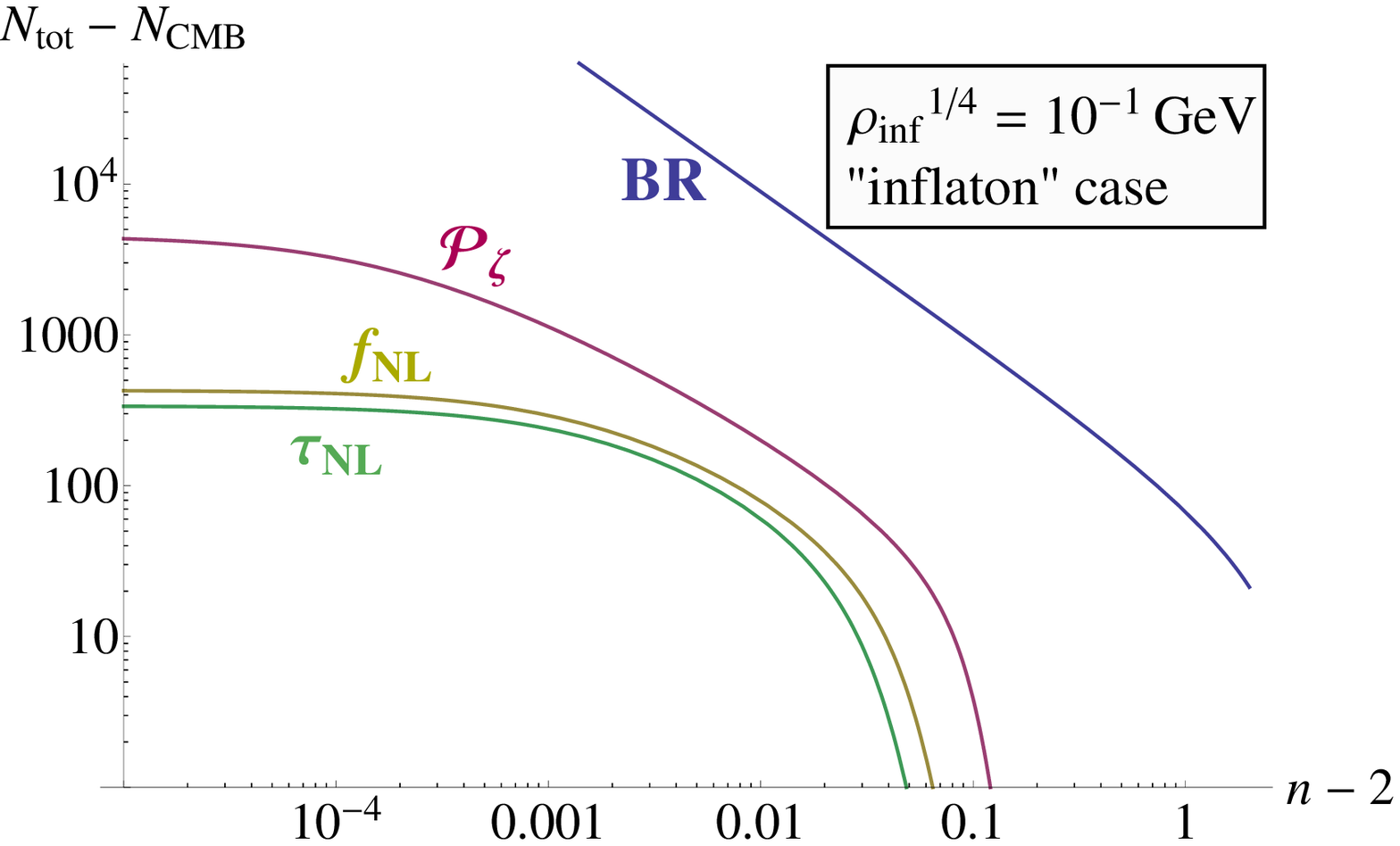}
 \caption
 {The upper limit of $N_{\rm tot}-N_\CMB$ for $n\ge2$ when
 inflaton generates the observed curvature perturbation. 
 The horizontal axis is $n-2$ and the vertical axis is 
 $N_\tot-N_\CMB$.
 The inflation energy scale is set as 
 $\rho_\inf^{1/4}=10^{16}$GeV (left panel) or  $10^{-1}$GeV (right panel).
 The blue line denotes the upper limit of $N_\tot -N_\CMB$ coming from the  back reaction condition, $\rho_\inf>\rho_\em$, while  the red, yellow and  green lines represent the upper limit from the induced 
 $\mathcal{P}_\zeta$, $f_{\rm NL}$ and $\tau_{\rm NL}$ from the electromagnetic field respectively.
In both panels, one can see that 
 the smaller the $N_\tot$ or $n-2$ is, the milder the constraints are.}
 \label{fig:NtNc-inf}
\end{figure}

In fig.~\ref{fig:NtNc-inf}, we plot the upper limit on $N_\tot -N_\CMB$
of the ``inflaton" case with changing $n$.
From these figures, we find that the constraint becomes more stringent as $n$ becomes larger. 
It is because the generated electric field becomes stronger for larger $n>2$
(see eq.~(\ref{E and B spectrum})) and thus
the induced curvature perturbation is amplified.
Aside from the back reaction constraint eq.~(\ref{NtNc from BR}), 
the upper limit from m-point correlator contains the factor
$\left(\frac{8}{3}c_n D_n(N_\CMB)\right)^{-m}$ in the argument of logarithm.
In case with $n= 2$, it reads
\begin{equation}
\left(
\frac{8}{3}c_n D_n(N_\CMB) \
\right)^{-m}
\xrightarrow{\scriptstyle n\to2} \
\left(600 \left(\frac{N_\CMB}{50}\right)
\right)^{-m}
\qquad
(m=2,3,4)
\label{hierarchy factor}
\end{equation} 
and it is even smaller for $n>2$.
Because of this factor, the higher $m$ is,
the more stringent the constraint is.
This behavior can be seen in fig.~\ref{fig:NtNc-inf} as the fact that
the constraint of $\tau_{\rm NL}$ is the tightest in the left panel.
Since low $\rho_\inf$ corresponds to low $N_\CMB$ as shown in eq. (\ref{NCMB}),
the hierarchy among the constraints
derived from $\mathcal{P}_\zeta, f_{\rm NL}$ and $\tau_{\rm NL}$ is
less significant 
as can be seen in the right panel of fig. \ref{fig:NtNc-inf} where we plot the upper limit of $N_\tot -N_\CMB$ for
$\rho_\inf^{1/4} = 10^{-1} {\rm GeV}$ case.
For $n=2$ case, the upper limit from $\tau_{\rm NL}$
can be obtained from eq.~(\ref{NtNc from taunl}) as
\begin{equation}
N_\tot-N_\CMB 
\ \lesssim\ 
17\times\left(\frac{N_\CMB}{50}\right)^{-4}
\left( \frac{\tau_{\rm NL}^\obs}{2800} \right),
\quad
(n=2,\ {\rm ``inflaton"\ case})
\,.
\end{equation}

\begin{figure}[htbp]
  \hspace{-2mm}
  \includegraphics[width=75mm]{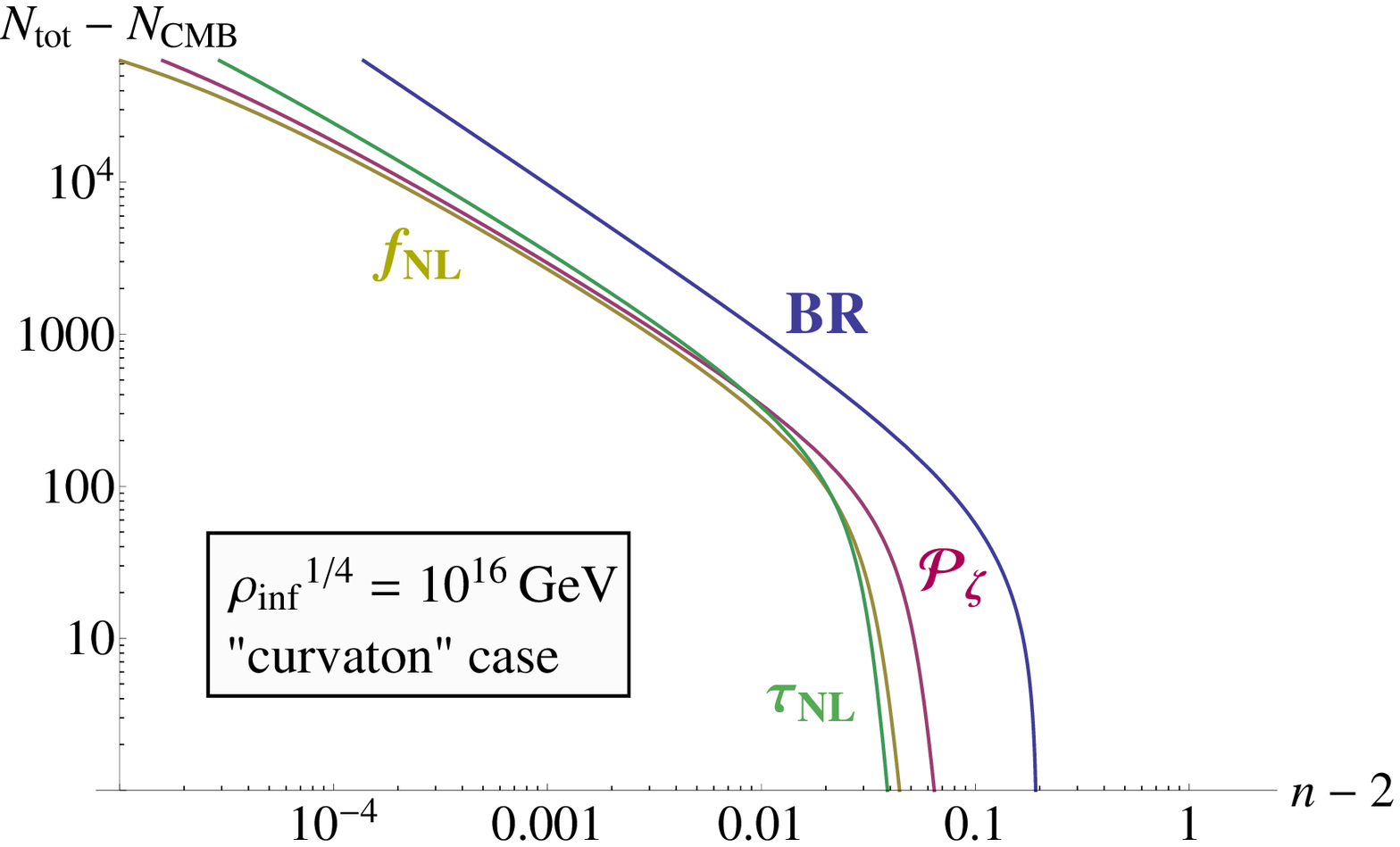}
  \hspace{5mm}
  \includegraphics[width=75mm]{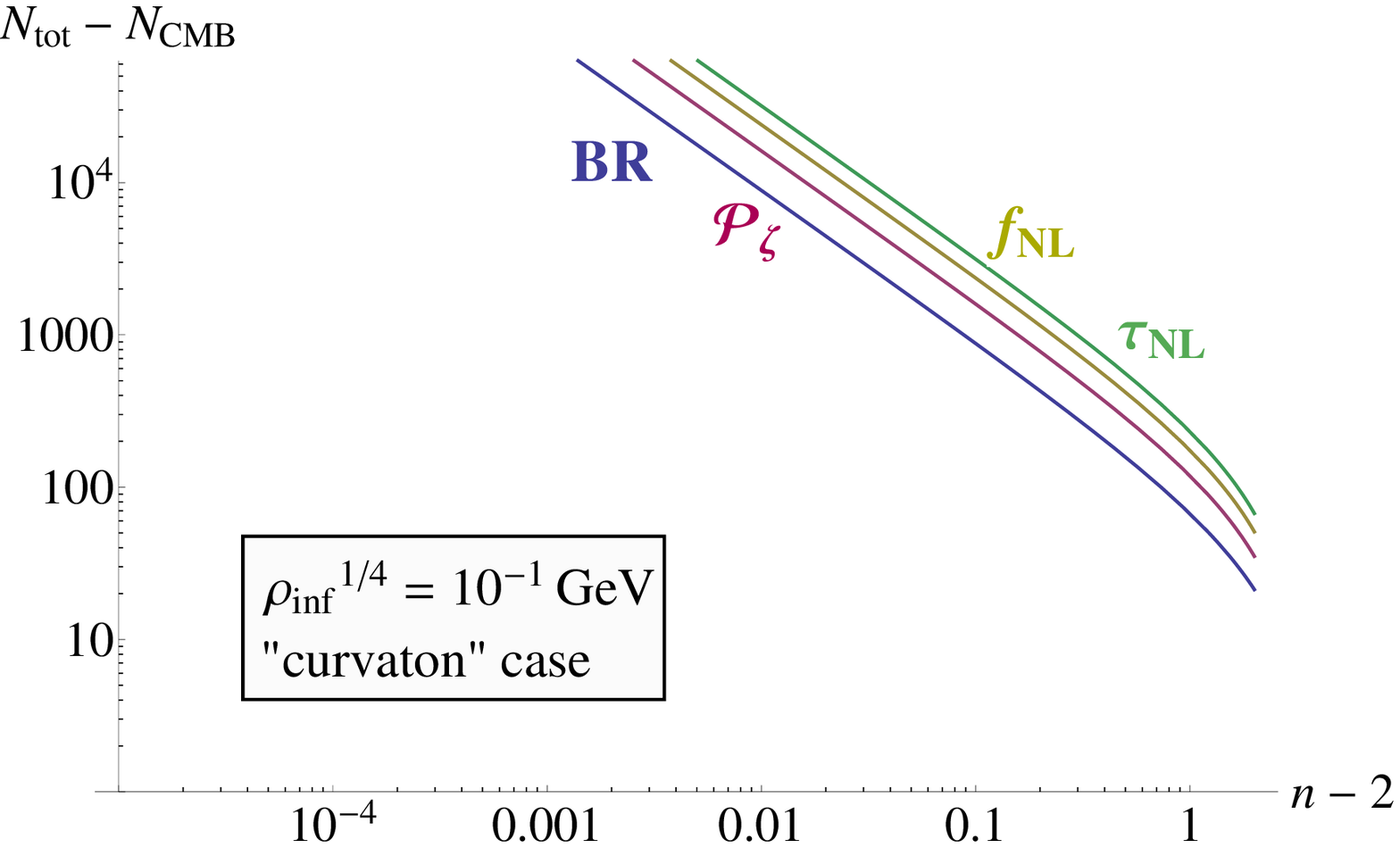}
 \caption
 {The upper limit of $N_{\rm tot}-N_{\CMB}$ for $n\ge2$ when
 the assumption that inflaton generates the observed curvature 
 perturbation is relaxed. 
 The horizontal axis is $n-2$ and the vertical axis is 
 $N_\tot-N_\CMB$.
 The inflation energy scale and slow-roll parameter are set as 
 $\rho_\inf^{1/4}=10^{16}$GeV (left panel) or $10^{-1}$GeV (right panel)
 and $\epsilon=10^{-2}$, respectively.
 The blue line denotes the upper limit of $N_\tot -N_\CMB$ coming from the  back reaction condition, $\rho_\inf>\rho_\em$, while  the red, yellow and  green line represent the upper limit from the induced 
 $\mathcal{P}_\zeta$, $f_{\rm NL}$ and $\tau_{\rm NL}$ from the electromagnetic field respectively. 
The back reaction constraint is unchanged from the ``inflaton"
case since it does not depend on $\epsilon$. But one can see the other
three constraints are much milder than those in fig.\ref{fig:NtNc-inf}.}
\label{fig:NtNc-curv}
\end{figure}

In fig.~\ref{fig:NtNc-curv}, we plot the upper limit on $N_\tot -N_\CMB$
of the ``curvaton" case by setting $\epsilon= 10^{-2}$.
In this figure, one can see that
the constraint is considerably milder than the
``inflaton" case.
It is interesting to note that the hierarchy among the four constraint
is inverted in the low $\rho_\inf$ plot (right panel).
In fact, the upper bound from the back reaction problem is most stringent
for $\rho_\inf^{1/4} \lesssim 10^{15}$GeV.
Except for eq.~(\ref{NtNc from BR}), 
the upper limit from m-point correlator contains the factor
$\mathcal{P}_\zeta^{m-1}/\mathcal{P}_\inf^{m}$ in the argument of logarithm.
Although it reads $\mathcal{P}_\zeta^{-1}$ in the ``inflaton" case,
in the ``curvaton" case it yields an extra factor,
\begin{equation}
\left(
\frac{\mathcal{P}_\zeta}{\mathcal{P}_\inf}
\right)^m
\simeq
\left(
18\times
\left(\frac{\epsilon}{0.01}
\right)
\left(\frac{(10^{16}\GeV)^4}{\rho_\inf}
\right)
\right)^m
\qquad
(m=2,3,4).
\end{equation}
Therefore the constraints from higher correlator
substantially relaxed especially in low $\rho_\inf$ region.
At $\rho_\inf^{1/4}\simeq 10^{16}$GeV, this factor compensates the factor
of eq.~(\ref{hierarchy factor}) and three constraints from
$\mathcal{P}_\zeta, f_{\rm NL}$ and $\tau_{\rm NL}$ are almost degenerate (see the left panel).
They are coincident with the back reaction constraint
at $\rho_\inf^{1/4}\simeq 10^{15}$GeV.
Thus the back reaction bound is the most stringent for 
$\rho_\inf^{1/4} \lesssim 10^{15}$GeV.

%

One can understand why the ``curvaton" case with 
$\epsilon=10^{-2}$
gives much milder bound than
the ``inflaton" case as follows. 
From eq.~(\ref{abb form P})-(\ref{abb form tau}), one can find 
$\mathcal{P}_\zeta^\em, f^\em_{\rm NL}$ and $\tau^\em_{\rm NL}$
are increasing function of $\rho_\inf$ and decreasing function
of $\epsilon$.
Thus one way of relaxing the upper limit is to increase $\epsilon$. But
$\epsilon$ can not vary freely in the ``inflaton" case because $\epsilon$
is determined by $\rho_\inf$ as
\begin{equation}
\epsilon =5.5\times 10^{-4} \left(\frac{\rho_\inf}{(10^{16}\GeV)^4}\right).
\label{fixed epsilon}
\end{equation}
Therefore the ``curvaton" case does not always put milder constraint
than the ``inflaton" case but it does only when $\epsilon$ is larger
than eq.~(\ref{fixed epsilon}).

\subsection{Constraint on the inflation energy scale $\rho_\inf$}

If we change the set of input parameters from $\{n, \epsilon, \rho_\inf\}$ into $\{n, \epsilon, N_\tot\},$ we can constrain $\rho_\inf$ instead of 
$N_\tot - N_\CMB$. Although eq.~(\ref{backreaction rho upper bound})
gives the upper limit of $\rho_\inf$ explicitly,
we have to numerically calculate 
the bounds from $\mathcal{P}_\zeta, f_{\rm NL}$ and $\tau_{\rm NL}$.
Provided that $N_\tot > \frac{3}{2}N_\CMB$,
one can show that the constraints
from $\mathcal{P}_\zeta, f_{\rm NL}$ and $\tau_{\rm NL}$
give  upper limits on $\rho_\inf$.
\footnote{
One can find the condition when $\mathcal{P}_\zeta^\em, f_{\rm NL}^\em$
and $\tau_{\rm NL}^\em$ are increasing function of $\rho_\inf$
by differentiating them with respect to $\rho_\inf$
and looking at their sign. It can be shown the conditions are 
$N_\tot > \frac{m+1}{m}N_\CMB, (m=2,3,4)$ in the ``inflaton" case
while the conditions are far milder in the ``curvaton" case.
}
Thus we adopt $N_\tot =100, 300$ and 1000 as the fiducial values.
Note
the energy scale of inflation is naively
restricted by
the indirect observation of gravitational wave
and the big bang nucleosynthesis as
\begin{equation}
10^{-1}\GeV 
\ \lesssim\ 
\rho_\inf^{1/4}
\ \lesssim \ 
10^{16}\GeV\,, 
\end{equation}
regardless of the kinetic coupling model.

\begin{figure}[htbp]
  \hspace{-2mm}
  \includegraphics[width=75mm]{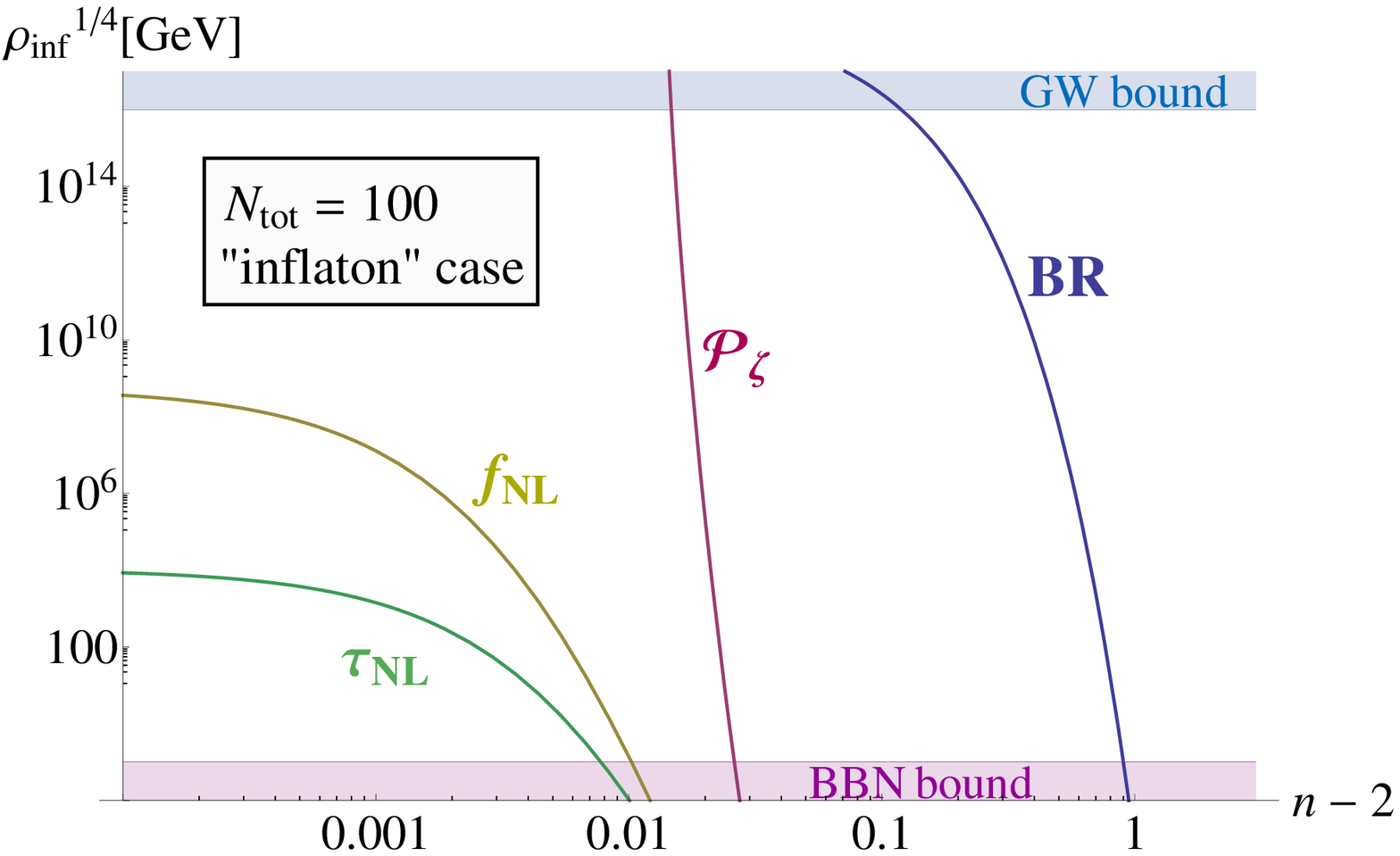}
  \hspace{5mm}
  \includegraphics[width=75mm]{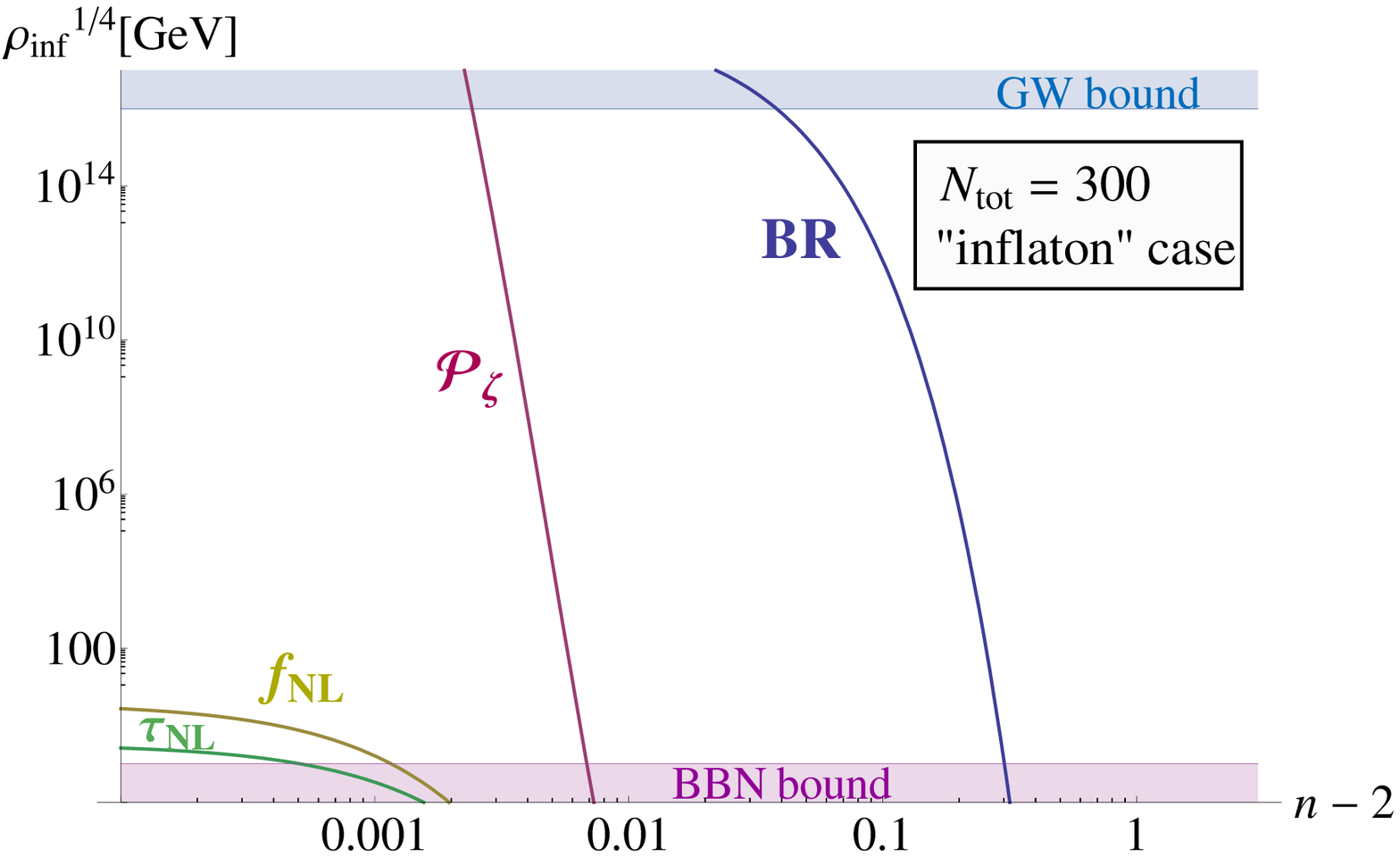}
  \\ 
  
  \hspace{-2mm}
  \includegraphics[width=75mm]{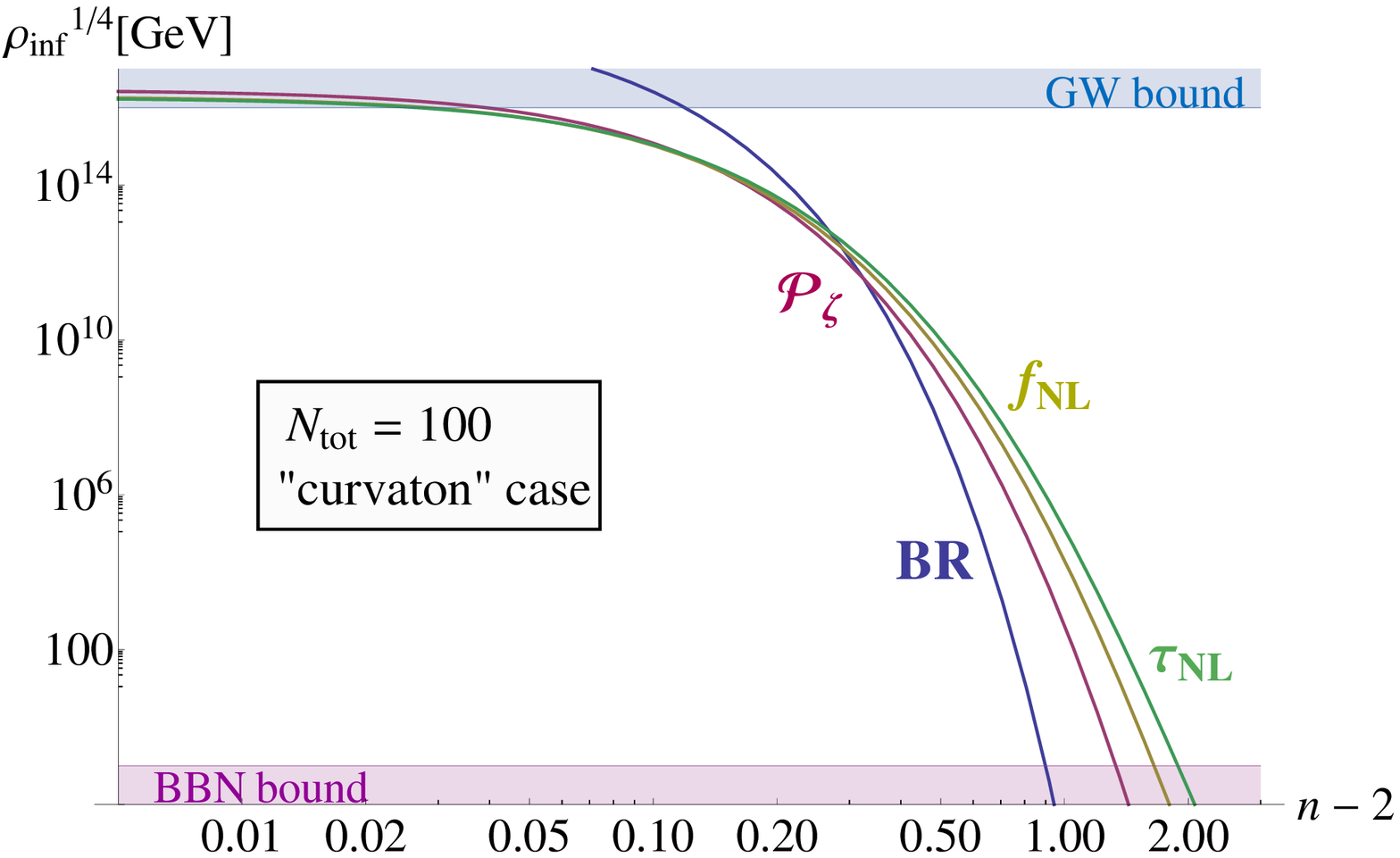}
  \hspace{5mm}
  \includegraphics[width=75mm]{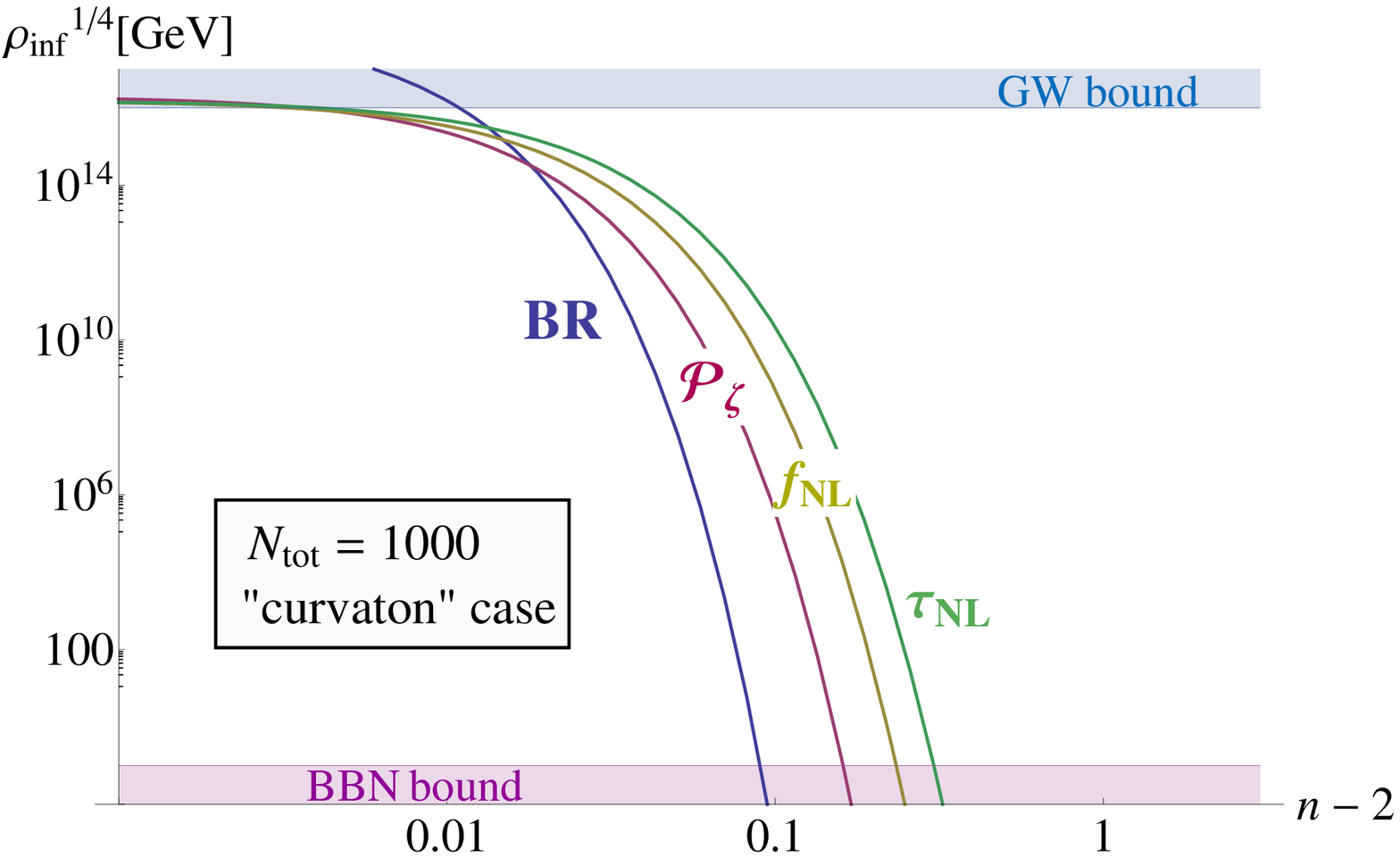}
 \caption
 {The upper limit of $\rho_\inf^{1/4}$ for $n\ge2$.
 The horizontal axis is $n-2$ and the vertical axis is 
 $\rho_\inf^{1/4}$ [GeV].
 In top two panels it is assumed that
 inflaton generates all observed curvature perturbation
 (``inflaton" case) while that assumption is relaxed and instead 
 $\epsilon = 10^{-2}$ is adopted in
 the bottom two panels (``curvaton" case). 
 The total duration of the electromagnetic field generation is set as 
 $N_\tot = 100$ (left panels), 300 (top right panel) or 1000 (bottom right  panel).
 The shaded regions represent the restriction from
 gravitational wave (blue) and big bang nucleosynthesis (red), respectively.
 }
 \label{fig:rho}
\end{figure}

In fig.~\ref{fig:rho}, we plot the upper limits on $\rho_\inf^{1/4}$.
The basic property of the constraint is unchanged from that on 
$N_\tot-N_\CMB$ because the origin of constraints is same.
Again, one can see that the larger $n$ is, the tighter the
constraints are. $\tau_{\rm NL}$ gives the most stringent bound
in the ``inflaton" case while the bound from the back reaction problem 
is the tightest in low energy region of the ``curvaton" case. 
In addition, now it is clear that the lower $\rho_\inf$ is,
the milder the constraints are. 
It is remarkable that $N_\tot\gtrsim300$ is excluded
in the ``inflaton" case. 
It is consistent with the right panel of fig.~\ref{fig:NtNc-inf}.
Even if $N_\tot<300$, $n$ and $\rho_\inf$ are severely restricted
in the ``inflaton" case.
On the other hand, the constraints in the ``curvaton" case are
much more moderate. Especially $\rho_\inf$ is free from 
a new restriction if $n$ is sufficiently small.
Furthermore, at low energy region, the tightest constraint is
given by the back reaction condition whose analytic formula is available.
Since in the right hand side of eq.~(\ref{backreaction rho upper bound})
the most important factor is $\exp[(2n-4)N_\tot]$, 
eq.~(\ref{backreaction rho upper bound}) can be approximated
by $n-2 \lesssim \ln(\Mpl^4/\rho_\inf)/2N_\tot$.
Then the largest allowed $n$ at $\rho_\inf^{1/4}=10^{-1}$GeV
is 
\begin{equation}
n-2\ \lesssim\ \frac{90}{N_\tot}
\,,\qquad
({\rm ``curvaton"\ case}\, 
).
\end{equation}
Since $N_\CMB$ is as small as $\approx 23$
at such low energy scale, $n$ can be larger than 4 in principle.
However, the resultant magnetic field strength at present is
depends on $\rho_\inf$ as $\mathcal{P}_B \propto \rho_\inf^{(n-1)/4}$
and thus a large $n$ does not necessarily lead to a strong magnetic field.

\subsection{Constraint on the strength of the magnetic field $B$}

\begin{figure}[htbp]
  \hspace{-2mm}
  \includegraphics[width=75mm]{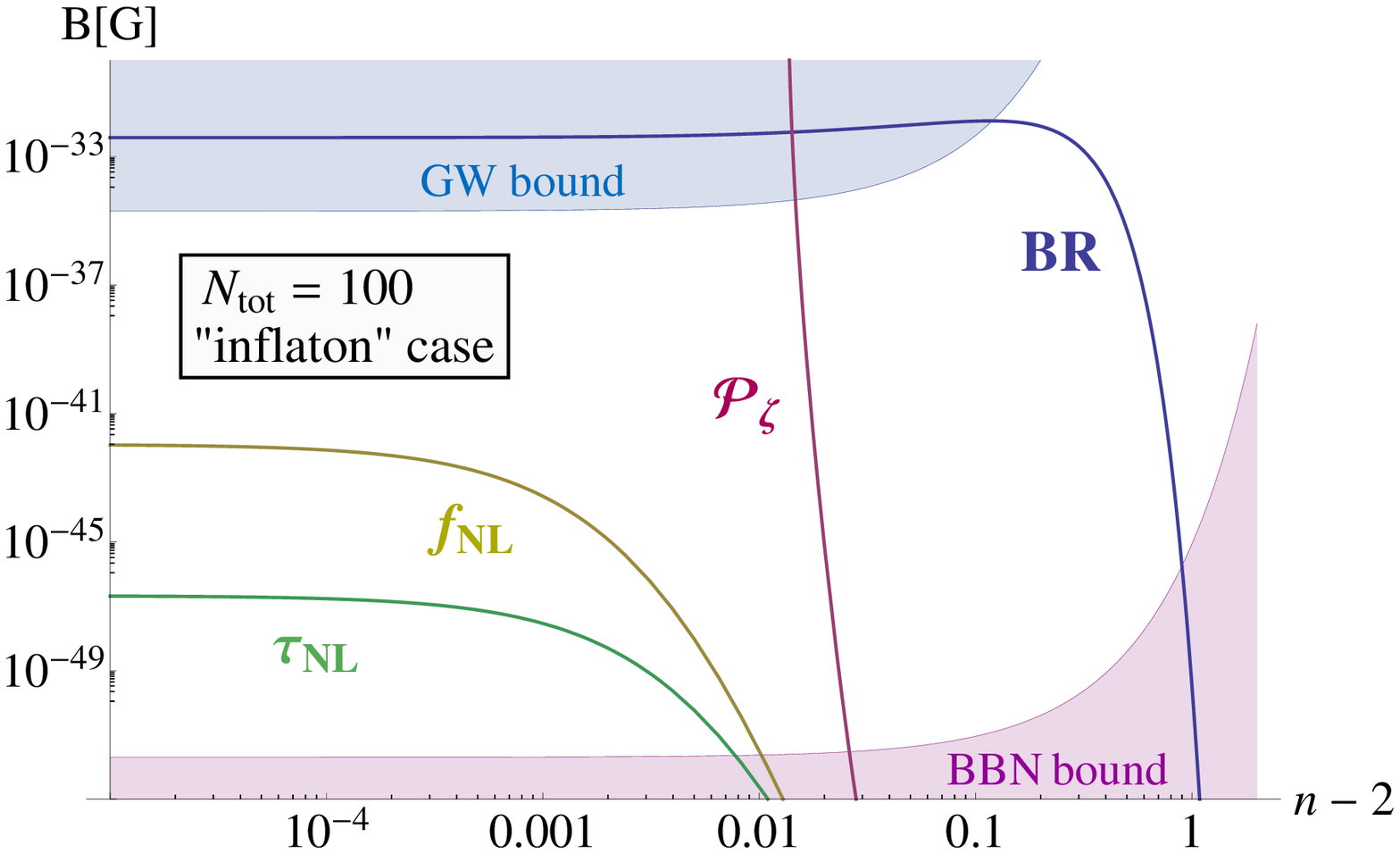}
  \hspace{5mm}
  \includegraphics[width=75mm]{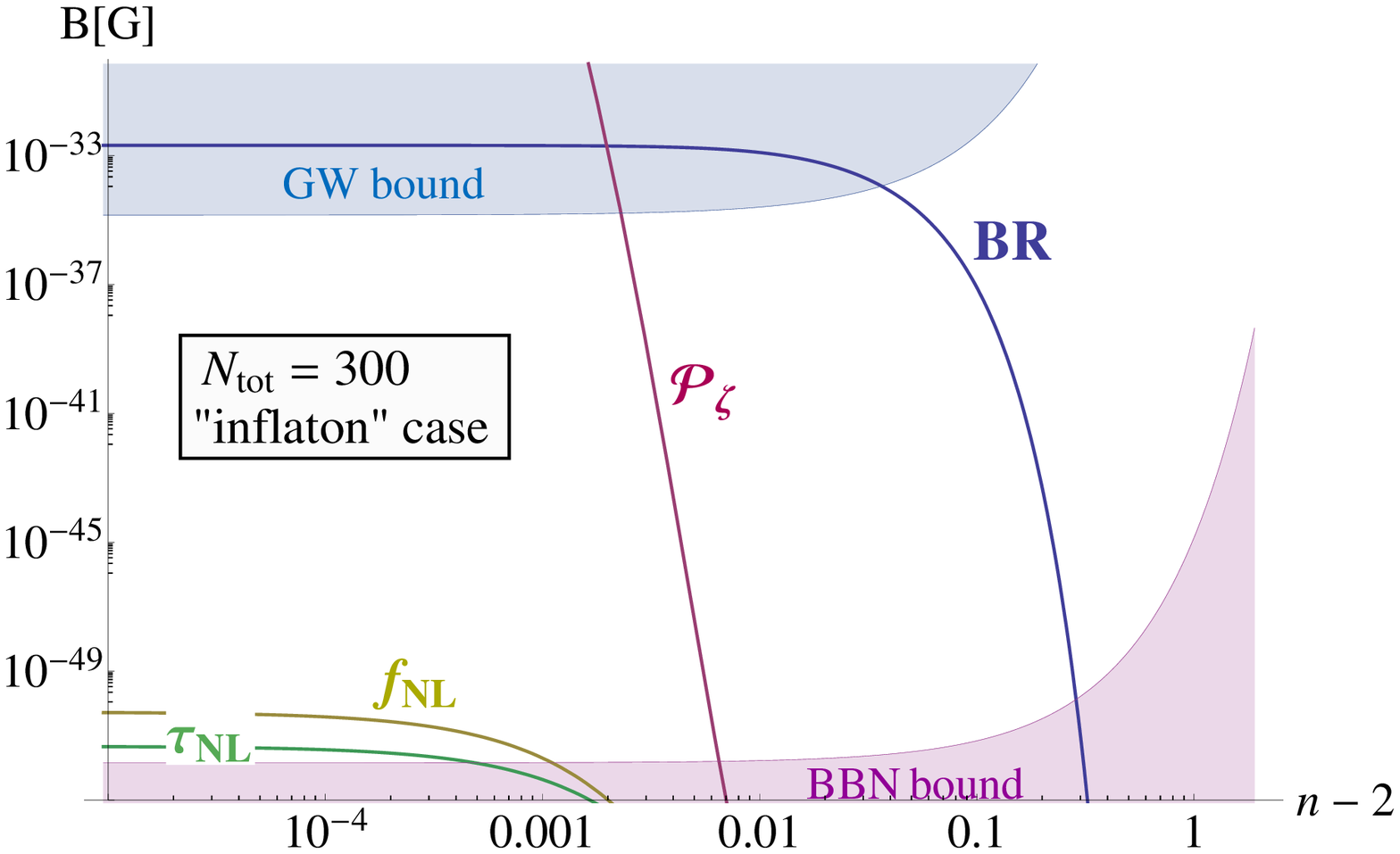}
  \\ 
  
  \hspace{-2mm}
  \includegraphics[width=75mm]{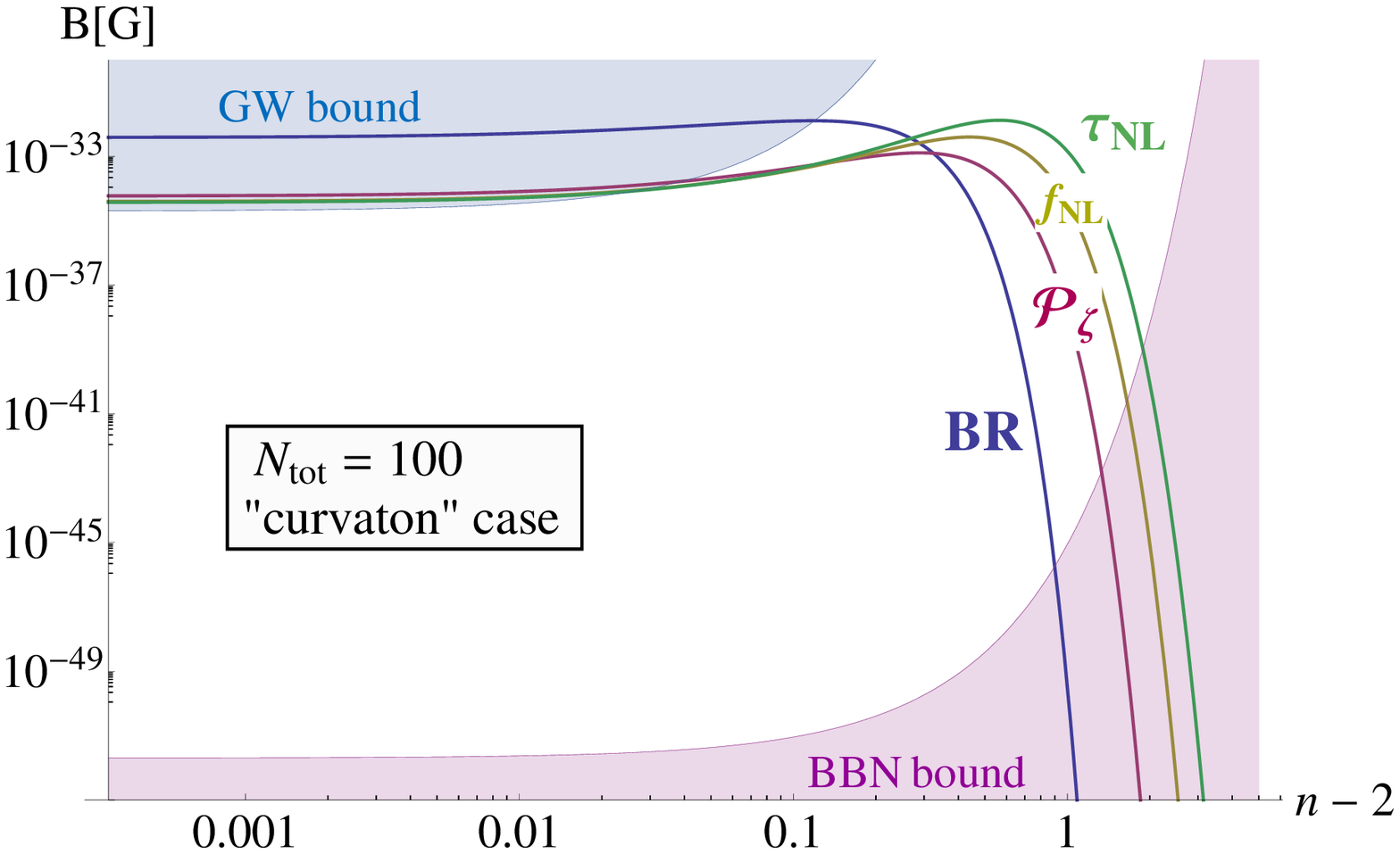}
  \hspace{5mm}
  \includegraphics[width=75mm]{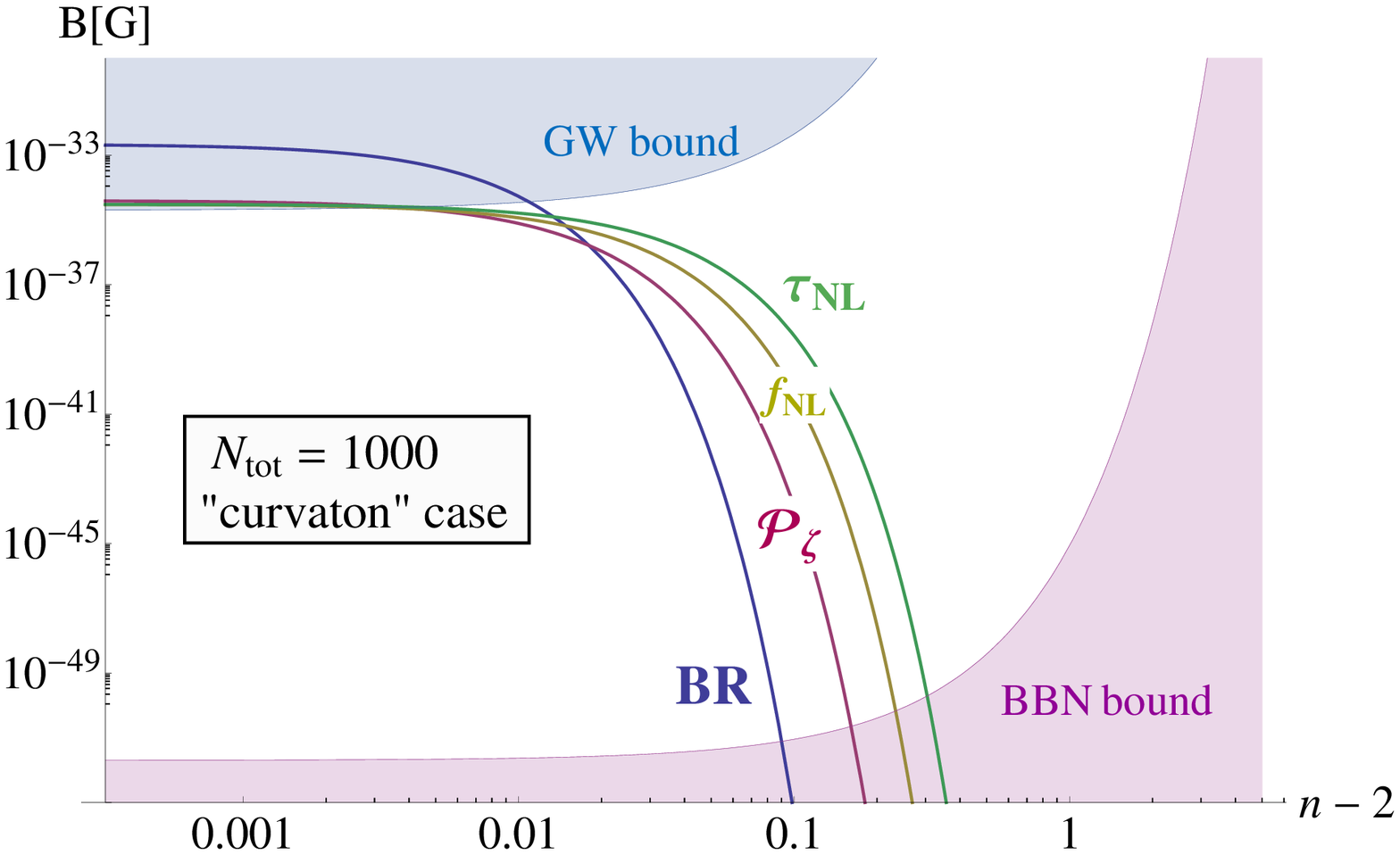}
 \caption
 {The upper limit of the current strength of the magnetic field
  for $n\ge2$.
 The horizontal axis is $n-2$ and the vertical axis is 
 $\mathcal{P}_B^{1/2}(\eta_{\rm now},1{\rm Mpc}^{-1})$ [G].
 In top two panels it is assumed that
 inflaton generates all observed curvature perturbation
 (``inflaton" case) while that assumption is relaxed and instead 
 $\epsilon = 10^{-2}$ is adopted in
 the bottom two panels (``curvaton" case). 
 The total duration of the electromagnetic field generation is set as 
 $N_\tot = 100$ (left panels), 300 (top right panel) or 1000 (bottom right  panel).
 The shaded region represent the restriction from
 gravitational wave (blue) and big bang nucleosynthesis (red), respectively.
 }
 \label{fig:B}
\end{figure}

In terms of magnetogenesis, it is interesting to put the upper limit
on the present strength of the magnetic field, $\mathcal{P}_B(\eta_{\rm now},k)$.
Combined with eq.~(\ref{current B}), the upper limits on $\rho_\inf$ which we obtain in the previous subsection by numerical calculations can be
converted into the upper limits on $\mathcal{P}_B(\eta_{\rm now},k)$.
Those limits are shown in fig.~\ref{fig:B}.

It is known that the strength of magnetic field generated is
kinetic coupling model has been already bounded above due to the back reaction problem
and its present value can not exceed $10^{-32}$G for $N_\tot=70$ and
$k=1{\rm Mpc}^{-1}$~\cite{Demozzi:2009fu}.
But it turns out that the upper limit is $10^{-47}$G due to 
the constraint from $\tau_{\rm NL}$ in the ``inflaton" case
(see the top left panel of fig.~\ref{fig:B}).
If $N_\tot$ is larger, the constraint becomes even severer.
On the other hand, in the ``curvaton" case, the strongest value
of magnetic field in the allowed region is smaller by only a few orders
of magnitude than that without the curvature perturbation constraints.

\section{Conclusion}
\label{Conclusion}

The kinetic coupling model (or $IFF$ model) has drawn attention as
both a magnetogenesis model and a generation mechanism of 
the curvature perturbation and non-gaussianities. 
Although it is known that the back reaction problem (BR) and the strong
coupling problem restrict this model from generating the magnetic
field which is strong enough to explain the blazar observation
at present, the constraints from the curvature perturbation
induced by the electromagnetic fields during inflation
are not yet investigated adequately.

In this paper, we compute the curvature power spectrum $\mathcal{P}_\zeta$
and non-linear parameters $f_{\rm NL}^{\rm local}, \tau_{\rm NL}$
of the curvature perturbation induced by the electromagnetic fields
in the kinetic coupling model with $I\propto a^{-n}$ for $n\ge 2$.
Quite recently $\mathcal{P}_\zeta, f_{\rm NL}^{\rm local}$ and 
$\tau_{\rm NL}$ are precisely determined or constrained by
the Planck collaboration. Thus by using the Planck result,
we constrain the parameters of the kinetic coupling model and inflation.
We find that $\mathcal{P}_\zeta^\em, f_{\rm NL}^\em$ and $\tau_{\rm NL}^\em$
are given by the functions of four parameters
$\{ n, N_{\rm tot},\rho_\inf, \epsilon\}$ of the model and inflation
(see eq.~(\ref{em power spectrum}), (\ref{em fnl}) and (\ref{em taunl})).
Therefore when three parameters out of four are fixed,
the other one can be constrained by the observation.
Note in the case where a single slow-rolling inflaton is responsible for
all the observed curvature power spectrum,
which we call ``inflaton" case, the slow-roll parameter $\epsilon$ is determined
by inflation energy scale $\rho_\inf$.
On the other hand, if the other mechanism like curvaton or modulated reheating produces observed $\mathcal{P}_\zeta$, $\epsilon$ can be a free parameter.
For simplicity, this case is called ``curvaton" case while we do not 
specify any model.

In order to illustrate the constraints from 
the BR, $\mathcal{P}_\zeta^\em, f_{\rm NL}^\em$ 
and $\tau_{\rm NL}^\em$,
we show three kinds of plot which represent the upper limit of
$N_\tot - N_\CMB,\, \rho_\inf$ and $\mathcal{P}_B^{1/2}(\eta_{\rm now},1{\rm Mpc}^{-1})$ with respect to $n$, respectively.
The upper limits of the total e-folding number of magnetogenesis
before the CMB scale exits the horizon,
$N_\tot-N_\CMB$, can be expressed by analytical formula
as eq.~(\ref{NtNc from BR})-(\ref{NtNc from taunl}).
The upper limits of the inflation energy density, $\rho_\inf$,
need numerical calculations to be obtained and can be translated to
the upper limits of the present amplitude of the cosmic magnetic field at Mpc scale,
$\mathcal{P}_B^{1/2}(\eta_{\rm now},1{\rm Mpc}^{-1})$.
In general, all four constraints from the BR, $\mathcal{P}_\zeta^\em,
f_{\rm NL}^\em$ and $\tau_{\rm NL}^\em$ become tighter as $n\, (\ge 2)$ is larger.
It is simply because the strength of generated electromagnetic fields 
are amplified as $n\, (\ge 2)$ is larger.

In the ``inflaton" case, interestingly, $\tau_{\rm NL}$ gives the strongest limitation on parameters.
Even for $\rho_\inf^{1/4}=10^{-1}\GeV$ and $n=2$, the constraint from $\tau_{\rm NL}$
puts $N_\tot \lesssim 300$ and it becomes more stringent 
for higher $\rho_\inf$ or $n$. 
For $N_\tot=100$ and $n=2$, in turn, $\rho_\inf^{1/4}\lesssim 10^4\GeV$ is
required and $\rho_\inf$ should be even lower for larger $N_\tot$ or $n$.
As for the magnetic field strength, we find
the upper limit from $\tau_{\rm NL}$ is $\mathcal{P}_B^{1/2}\lesssim 10^{-47}$G at present Mpc scale for $N_\tot = 100$. It is $10^{-15}$ times lower than
the upper limit of the conventional BR condition.

In the ``curvaton" case, however, the constraints are more moderate
if the free parameter $\epsilon$ is larger than the ``inflaton" case.
For clarity we fix $\epsilon=10^{-2}$ and show the constraints from
$\mathcal{P}_\zeta^\em, f_{\rm NL}^\em$ and $\tau_{\rm NL}^\em$
are weaker than the BR constraint if $\rho_\inf$ is sufficiently small.
Thus even if the induced curvature perturbation is taken into account,
the resultant constraint is not dramatically changed 
from the conventional BR restriction in the low $\rho_\inf$ region.
In fact, one can see in fig.~\ref{fig:B}
that the constraint on $\mathcal{P}_B$ at present Mpc scale
becomes tighter only by $\mathcal{O}(10^{-1})$
than that given solely by the BR. 

Aside from the constraints, we find the general relationship between
$f_{\rm NL}^\em$ and $\tau_{\rm NL}^\em$ in eq.~(\ref{f and t relationship}).
According to it, even if $f_{\rm NL}\sim \mathcal{O}(1)$ which is too small to be observed by the Planck satellite,
the kinetic coupling model can compatibly produce 
detectable $\tau_{\rm NL}\gtrsim 560$~\cite{Kogo:2006kh}.
In addition, it is expected that this model generates much higher correlators
of the curvature perturbation.
Thus it is also interesting to investigate the higher order correlators
both in theoretical and observational sides.
Furthermore, we use the averaging over the direction of $\hat{\bm{k}}_i$
for $f_{\rm NL}$ and $\tau_{\rm NL}$.
It should be interesting to consider the direction dependence
of $\tau_{\rm NL}$ as well as $f_{\rm NL}$.


\acknowledgments

We would like to thank Jun'ichi Yokoyama for useful comments. 
This work was supported by the World Premier International
Research Center Initiative (WPI Initiative), MEXT, Japan. 
T.F. and S.Y. acknowledge the support by Grant-in-Aid for JSPS Fellows
No.248160 (TF) and No. 242775(SY).


\end{document}